\documentclass[10pt,journal]{IEEEtran}

\usepackage[ruled,vlined]{algorithm2e}

\usepackage{bm} 

\usepackage{epsfig}
\usepackage{amssymb} 
\usepackage[tbtags]{amsmath} 

\usepackage{latexsym}
\usepackage{euscript}
\usepackage{graphics,eepic,epic,psfrag}
\usepackage{enumitem}
\usepackage{xhfill}
\usepackage{relsize}
\usepackage{exscale}
\usepackage{stfloats}%

\usepackage[font = footnotesize]{caption}
\usepackage{subcaption}
\usepackage{graphicx}
\usepackage{cite}  
\usepackage{xpatch}
\usepackage{algorithmicx}
\usepackage{comment}

\newcommand\numberthis{\addtocounter{equation}{1}\tag{\theequation}}
\newcommand{\upperRomannumeral}[1]{\uppercase\expandafter{\romannumeral#1}}
\newcommand{\lowerromannumeral}[1]{\romannumeral#1\relax}
\DeclareMathOperator{\tr}{tr}

\begin{document}
%
\title{Localization with Reconfigurable Intelligent Surface: An Active Sensing Approach}
%
%
\author{Zhongze~Zhang,~\IEEEmembership{Student Member,~IEEE,}
        Tao~Jiang,~\IEEEmembership{Student Member,~IEEE,}
        and~Wei~Yu,~\IEEEmembership{Fellow,~IEEE}
\thanks{The authors are with The Edward S. Rogers Sr. Department
of Electrical and Computer Engineering, University of Toronto, Toronto,
ON M5S3G4, Canada. E-mails: ufo.zhang@mail.utoronto.ca, 
taoca.jiang@mail.utoronto.ca, weiyu@ece.utoronto.ca.
The materials in this paper have been presented in part at the IEEE International Conference on Communications (ICC), 2023 \cite{active2023icc}.
This work is supported by Huawei Technologies Canada Ltd. Co. 
}
}

\maketitle

\begin{abstract} 
This paper addresses an uplink localization problem in which a base station (BS) aims to locate a remote user with the help of reconfigurable intelligent surfaces (RISs). We propose a strategy in which the user transmits pilots sequentially and the BS adaptively adjusts the sensing vectors, including the BS beamforming vector and multiple RIS reflection coefficients based on the observations already made, to eventually produce an estimated user position. This is a challenging active sensing problem for which finding an optimal solution involves searching through a complicated functional space whose dimension increases with the number of measurements. We show that the long short-term memory (LSTM) network can be used to exploit the latent temporal correlation between measurements to automatically construct scalable state vectors. Subsequently, the state vector is mapped to the sensing vectors for the next time frame via a deep neural network (DNN). A final DNN is used to map the state vector to the estimated user position. Numerical result illustrates the advantage of the active sensing design as compared to non-active sensing methods. The proposed solution produces interpretable results and is generalizable in the number of sensing stages.  Remarkably, we show that a network with one BS and multiple RISs can outperform a comparable setting with multiple BSs.

\end{abstract}

\begin{IEEEkeywords}
Active sensing, deep learning, localization, reconfigurable intelligent surface, recurrent neural network, long short-term memory (LSTM).
\end{IEEEkeywords}

%
\IEEEpeerreviewmaketitle

\section{Introduction}
%
%
%
%

\IEEEPARstart{M}{}odern 
technologies such as traffic planning, autonomous vehicles, and robotic navigation depend heavily on the availability of location data 
\cite{6924849,6gwhitepaper}. 
Radio frequency (RF) localization often relies on
metrics such as time-of-arrival (ToA), received signal strength (RSS),
angle-of-arrival (AoA), or angle-of-departure (AoD) for localization.  
For example, many traditional RF
positioning algorithms are developed based on the RSS values received at the
multiple base stations (BSs), which represent the power level of the received signal 
after channel attenuation. 
However, RSS-based localization 
are often subject to large errors, especially in rich multipath environments
\cite{Palaskar2014WiFiIP,4796924}. Localization performance can be improved with time-of-flight measurements, 
but these measurements require
accurate synchronization at different anchor points and hence the extra overhead.
Another effective way to improve the
localization accuracy is to utilize directional information that provides the geometric relation between the transceivers, e.g., by estimating
the AoA using large antenna arrays at the BS \cite{aoa_localization,aoa_localization2}, 
or by adding more anchor points, i.e., BSs or access points (AP)
\cite{nguyen}, but these options add to the infrastructure cost. 

This paper explores the use of reconfigurable intelligent surface (RIS) as a cost-effective
alternative to large antenna arrays for collecting directional information.
RIS is a reflector with a planar surface consisting of a large number of
passive elements, with each element capable of altering the phase of the incident electromagnetic wave with very low power consumption\cite{bible,
basar2019wireless, 4fuliu,coding_metasurface }. 
The technology is envisioned as one of the enabling technologies for 6G localization due to its ability to:
\lowerromannumeral{1})
establish a reliable reflected link when the direct path between the transceivers is weak or blocked, and \lowerromannumeral{2})
effectively provide additional anchor points without requiring additional RF chains  \cite{6gwhitepaper,panpan}. 
Many recent works have considered RIS-assisted localization both in terms of 
theoretical bounds and practical algorithms \cite{hejiguang,liu2020reconfigurable,siso, nearfield,mimoofdm,Ubiquitous,Ubiquitous2,wang2021joint,elzanaty2020reconfigurable, combination, multiriscrlb,nguyen}.





This paper considers an uplink localization problem with multiple RISs,
where a single user repeatedly transmits pilot symbols and the
BS receives the pilots through the reflection by the RISs and determines the location
of the user based on the received pilots. 
Specifically, we investigate a scenario in which the BS can control 
the reflection coefficients at the RISs and tackle the problem of designing the RIS 
configurations and BS beamforming vector in an \emph{active} manner to
minimize localization error. By active sensing, we mean that the RISs
reflection coefficients and the BS beamforming vector are sequentially designed as a function of previous
measurements. As a result, the BS and the RISs can be used to focus the beams progressively to
locate the user over time as more measurements become available. 

Active sensing can significantly improve the localization accuracy, but the design of active RIS configuration and BS beamformer is also a challenging problem, because finding the optimal solution requires navigating a complicated functional landscape over the sequence of observations. Since the number of measurements increases over time, a particularly important issue is how to find a scalable solution, which this paper seeks to address.
To make the problem tractable, \cite{hejiguang} proposes a hierarchical codebook for the RIS that enables adaptive bisection search over the angular space in a 2D localization setting. 
However, a hierarchical codebook is not necessarily the best approach, especially at low signal-to-noise ratio (SNR)
\cite{aoa_adaptive}. 
In another work on active sensing with RIS \cite{liu2020reconfigurable}, 
the authors use gradient descent method to optimize RIS reflection coefficients
with the goal of minimizing the Cram\'er-Rao lower bound (CRLB) in each step. 
However, this is a greedy policy and further the CRLB is not necessarily tight, so it may not be a good metric for designing the BS beamforming vector and RIS reflection coefficients.

This paper proposes a learning-based design of active sensing vectors (including the BS beamforming vector and multiple RIS reflection coefficients) for localization, where the sequence of sensing vectors is designed as a function of the measurements already made so far. 
To address the key design challenge that as the dimension of historical observation increases with time, the functional landscape of designing the next sensing vectors becomes increasingly complex, we 
make use of recurrent neural network (RNN) with long short-term memory (LSTM) units to automatically construct an information vector of fixed dimension
from existing measurements and to extract temporal features
and long-term dependencies from a sequence of temporal
input. 
This is a generalization of the neural network architecture proposed in the previous work \cite{sohrabi2021active}, which deals with beam alignment problem in mmWave channel and RIS-assisted network, to a distinct solution tailored to localization in a multi-RIS-assisted multiple-input single-output (MISO) network.

More specifically, 
to design the beamforming vector and the RISs jointly for high localization accuracy based on the existing measurements, we use a chain of LSTM cells corresponding to the measurement time frames, to exploit the latent temporal correlation between the existing measurements and to construct/update information vectors of fixed dimension (hidden state and cell state). 
At each time frame, the information vectors are updated based on the newly received measurement and historical measurements. The next set of RIS configurations and beamforming vector are then designed from the hidden state via a deep neural network in a codebook-free fashion. After all the measurements have been received, a final deep neural network is used to map the cell state to the estimated user position. 
Numerical result shows that in a single-RIS-assisted single-input single-output (SISO) network, the proposed algorithm can achieve lower localization error as compared to existing non-active sensing methods and outperforms conventional active sensing method i.e., Bayesian Cram\'er-Rao lower bound (BCRLB) based approach. Moreover, we show that a multi-RIS-assisted MISO network with hybrid active/passive anchors can perform better than the conventional trilateration system with multiple BS-based anchors. 
Remarkably, the proposed solution can produce interpretable beampatterns and is also generalizable.





\subsection{Related Work}

In the context of localization algorithms in RIS-assisted network, the design of RIS configuration to enhance localization accuracy is of great interest. 
The authors in \cite{nearfield} devise three classes of RIS profiles and evaluate the impact of each class on the 3D position error bound of the user equipment (UE).
Two of the three classes are heuristically designed to produce narrow/broad beams, 
and the remaining class consists of random RIS profiles. 
In \cite{mimoofdm}, the authors study a channel estimation and user positioning problem in multiple-input
multiple-output (MIMO) orthogonal frequency division multiplexing (OFDM) systems assisted by RISs. Four designs of RIS training coefficients, i.e., random, structured, grouping
and sparse patterns, are proposed to assist in estimating channel parameters. Subsequently, an AoA-based positioning scheme is used to recover UE position. 
In \cite{Ubiquitous,Ubiquitous2}, indoor localization accuracy is improved by designing a set of RIS profiles that enlarge the differences between the RSS values of adjacent locations. The set of RIS profiles is designed via local and global search methods.
The authors in \cite{wang2021joint} consider joint localization and communication optimization
problem and design the appropriate RIS profiles to improve
localization accuracy and transmission throughput. 
In near-field region, the authors in \cite{elzanaty2020reconfigurable} derive the CRLB for assessing the localization and orientation performance of synchronous and asynchronous
signalling schemes in an RIS-assisted localization system. A closed-form RIS phase profile design that maximizes the sum of the SNRs at each BS antenna while accounting for the spherical wavefront is proposed. 
In \cite{combination}, an RIS-aided downlink positioning problem from the Fisher information perspective is considered, where the discrete elements of the RIS are optimized combinatorially to improve the rank of the Fisher information matrix. The work echoes the intuition that the design of the RIS phase profile should concentrate the reflected
signal energy toward the UE.
The authors in \cite{multiriscrlb} consider a multiple-RIS-aided mmWave positioning system and derive the CRLB. The phase shifts at the RISs are optimized via the particle swarm optimization (PSO) algorithm with the goal of minimizing the sum position error bound and rotation error
bound. 
In \cite{nguyen}, the authors consider a fingerprinting localization problem enabled by an RIS. In particular, the authors employ machine learning to identify a subset of RIS profiles from a codebook to generate a more diverse set of fingerprints to improve matching accuracy. 

All of the above works have demonstrated the importance of designing/selecting an appropriate set of RIS configurations and the associated performance gain. However, \emph{active} design of RIS configuration, which could further improve the performance, has not been considered. 

The notion of active sensing for communications \cite{javidi} has been proposed in sequential beamforming design problem for the initial access phase in a mmWave environment with a single-path channel model \cite{aoa_adaptive_journal}, where 
a deep neural network is used to map the current estimate of AoA posterior distribution to the beamforming vector of the next measurement. 
The work is later generalized to multi-path channel model in \cite{sohrabi2021active} using an LSTM-based RNN. 
An independent work \cite{GRU} addresses the same multi-path sequential beamformer design problem using Gated Recurrent Units (GRUs). 
The LSTM-based RNN designed in \cite{sohrabi2021active} for beam alignment also finds application in beam tracking problem as shown in \cite{hantracking}, where the RIS is actively reconfigured to track the movement of the user over time. 
These works do not address active sensing for localization using RIS.



\subsection{Main Contribution}



This paper proposes a learning-based localization strategy, where an adaptive sequence of sensing vectors is designed based on the historical measurements of the environment to improve localization accuracy. 
This is accomplished by adopting a data-driven approach to optimize the mapping from existing measurements to the sensing configuration in the next sensing stage to enhance localization in a codebook-free fashion, in recognition of the rich UE positional information contained in the pilot symbols.
In particular, we use 
the LSTM-based network, for its ability to capture the temporal relationship between different measurements over a long period, and its robust performance against exploding/vanishing gradient problem \cite{LSTM}. 
At each measurement time frame, an LSTM cell accepts new measurement of the environments, i.e., received pilots, and automatically uses the newly
received measurement (along with historical measurements) to update a hidden state vector of fixed dimension. 
The hidden state vector is subsequently updated via multiple layers until the vector obtains the correct representation of information to design the sensing vector for the next time frame. 
After multiple measurements, the cell state of the final LSTM cell is passed through another fully connected neural network to obtain
the estimated UE position. 
The main novelties of the proposed solution are as follows:
\begin{enumerate}
    \item An active sensing based approach is proposed to localize the UE in a multi-RIS-assisted MISO network, where the beamforming vector at the BS and reflection coefficients at the RISs are sequentially designed based on previous measurements in a codebook-free fashion to enable better localization accuracy. 
    
    \item An RNN with a chain of LSTM cells, corresponding to the sequence of sensing stages, is proposed to map the existing received pilots to the next reflection coefficients at the RISs and the beamforming vector at the BS. By carefully designing the loss function, the LSTM network can be trained to be generalizable in the number of sensing stages. 
    
    \item The proposed learning approach produces interpretable results, showing that the beampattern reflected off the RIS learns to gradually focus towards the UE to enhance the SNR of the received pilots for localization. 
    
    
    

\end{enumerate}

Simulation results show that the proposed learning-based active RIS profile design achieves lower localization error as compared to the existing non-active RIS profile design and outperforms conventional active sensing strategy, i.e., the BCRLB-based method. 
When generalized to a multi-RIS system, simulation results show that the proposed learning-based adaptive approach 
can produce a higher localization accuracy than a conventional triangulation-based multi-BS system. This shows the feasibility and the advantage of a hybrid active/passive RIS system over the conventional multi-anchor-based localization scheme.

The conference version of this paper \cite{active2023icc} only considers active sensing with a single RIS in a SISO network. This
journal paper extends to active sensing with multi-RIS in a MISO network.

\subsection{ Organization of the Paper and Notations}
The remaining paper is organized as follows. Section \ref{sec.background} introduces the system model, pilot transmission scheme, and problem formulation. Section 
 \ref{sec.rnn} describes the RNN architecture. Section \ref{sec.crlb} discusses the BCRLB-based active sensing approach. Numerical results are provided in Section \ref{sec.numerical_irs_single} and Section \ref{sec.numerical_irs_multi}. The paper concludes with Section \ref{sec.conclusions}.

$\emph{Notations}$: We use $a$, $\bm{a}$, and $\bm{A}$ to denote scalar, vector, and matrix respectively, $\bm{A}^\top$ and $\bm{A}^{\sf H}$ to denote transpose and Hermitian, $|\cdot|$ to denote modulus, 
$[\bm{a}]_j$ to denote the $j$-th element of the vector $\bm{a}$,
$[\bm{A}]_{i,j}$ to denote the $(i,j)$-th element of the matrix $\bm{A}$,
and diag($\bm{a}$) to denote the diagonal matrix with the entries of $\bm{a}$ on the diagonal. We use $\circ$ to denote the element-wise product. We use $\tr(\cdot)$ to denote the trace operation on a square matrix. 
We use $\mathcal{R}(\cdot)$ and $\mathcal{I}(\cdot)$ to denote the real and imaginary components of a complex value; 
$\mathcal{C}\mathcal{N}(\cdot,\cdot)$ to denote a complex Gaussian distribution; $\mathbb{E}(\cdot)$ to denote the expectation of a random variable. We use ${\rm mod}(\cdot)$ to denote the modulo division and $\lfloor \cdot \rfloor$ to denote the floor function.

\section{System Model and Problem Formulation}
\label{sec.background}

\subsection{System Model}

\begin{figure}[t]
\centering
\includegraphics[width=\columnwidth]{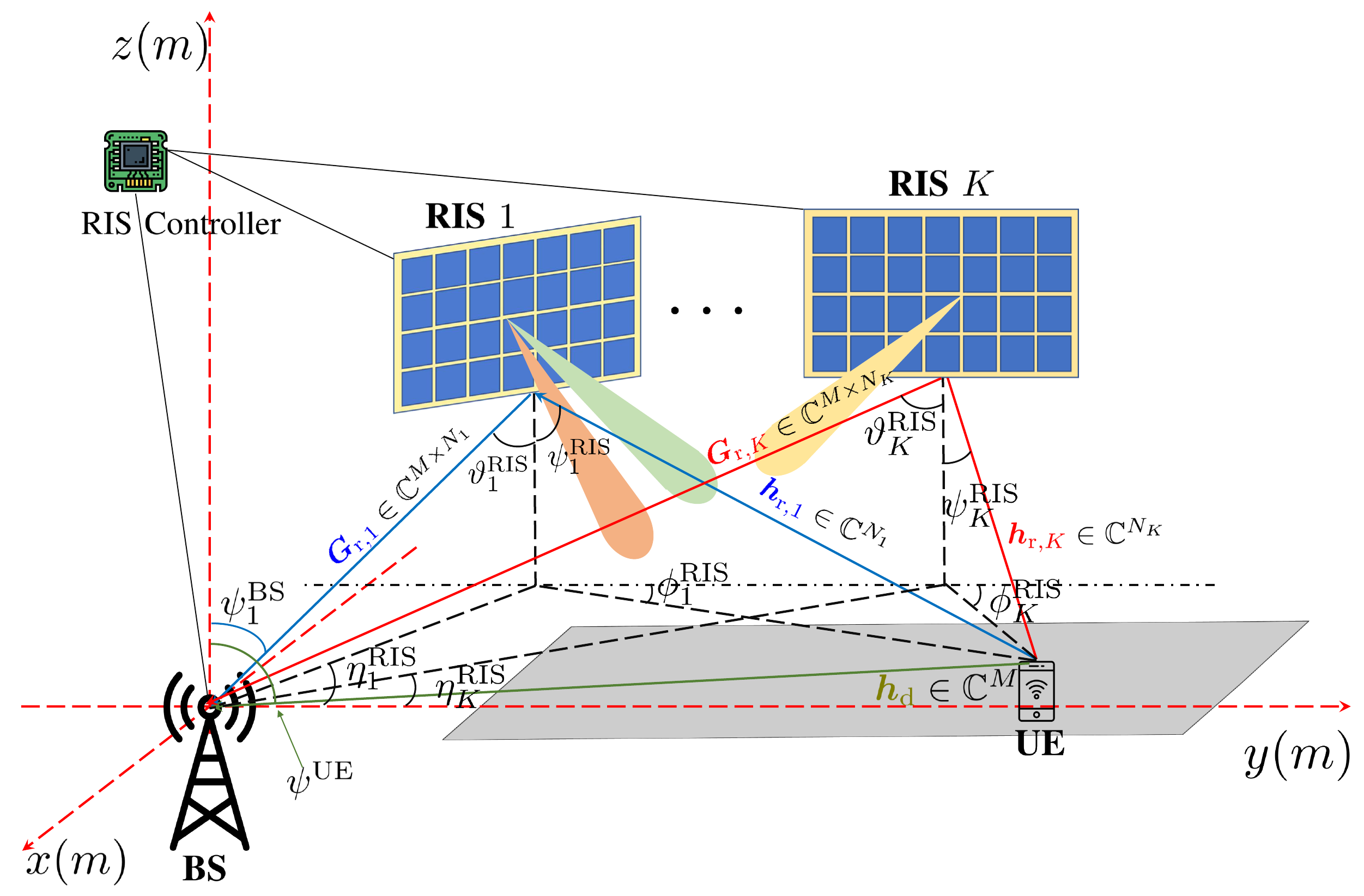}%
\caption{Multi-RIS MISO network.}%
\label{fig.setup_intro}%
\vspace{-0.5em}
\end{figure}

This paper considers a localization problem in an RIS-assisted system where a multi-antenna BS aims to estimate the location of a single-antenna UE with the aid of $K$ planar RISs. 
The BS and RISs are placed at the edge of a service area, and the user is located inside the serving area as in Fig.~\ref{fig.setup_intro}. 

The reflection coefficients of the $K$ RISs are controlled by an RIS controller which receives control signals from the BS. Let $N_k$ be the number of reflection coefficients at the $k$-th RIS. The reflection coefficients of the $k$-th RIS are denoted as
\begin{equation}
    \bm{\theta}_k = [e^{j\delta_1},  e^{j\delta_2}, \cdots , e^{j\delta_{N_k}}]^\top\in\mathbb{C}^{N_k},
\end{equation}
with $\delta_n \in [0,2\pi)$ as the phase shift of the $n$-th element.
The BS is equipped with $M$ antennas and a single RF chain. 
Let $\bm{w}\in\mathbb{C}^{M}$ denote the analog beamforming vector whose entries satisfy a constant modulus constraint, i.e., 
$[\bm{w}]_m = e^{j \omega_m}$ with $\omega_m \in [0,2\pi)$. 



We adopt a \emph{block-fading} model in which the channels are assumed to be constant within a coherence period, then change independently in subsequent coherence periods. 
As shown in Fig.~\ref{fig.setup_intro}, $\bm{h}_{\rm{d}}\in\mathbb{C}^M$ denotes the direct channel from the UE to the BS, $\bm{h}_{{\rm{r}},k}\in\mathbb{C}^{N_k}$ denotes the reflection channel from the UE to the $k$-th RIS, and $\bm{G}_{{\rm{r}},k} \in\mathbb{C}^{M \times N_k }$ denotes the channel from the $k$-th RIS to the BS. 
We ignore the multiple reflections across the multiple RISs due to the significant path loss of such reflections. 
We further assume that the reflection channel $\bm{h}_{{\rm r},k},\bm{G}_{{\rm r},k}$ and the direct channel $\bm{h}_{\rm d}$ follow Rician fading model:
\begin{subequations} 
\begin{align}
\bm{h}_{{\rm r},k} =\; & \kappa_k\left(\sqrt{\dfrac{\epsilon}{1+\epsilon}} \bm{\tilde{h}}_{{\rm r},k}^{\rm LOS} + \sqrt{\dfrac{1}{1+\epsilon}}\bm{\tilde{h}}_{{\rm r},k}^{\rm NLOS}\right),\\
\bm{G}_{{\rm r},k} =\; & \xi_k\left(\sqrt{\dfrac{\epsilon}{1+\epsilon}}\bm{\tilde{G}}_{{\rm r},k}^{\rm LOS} + \sqrt{\dfrac{1}{1+\epsilon}}\bm{\tilde{G}}_{{\rm r},k}^{\rm NLOS}\right),\\
\bm{h}_{\rm d}  =\;& \rho\left(\sqrt{\dfrac{\epsilon}{1+\epsilon}} \bm{\tilde{h}}_{{\rm d}}^{\rm LOS} + \sqrt{\dfrac{1}{1+\epsilon}}\bm{\tilde{h}}_{{\rm d}}^{\rm NLOS}\right)\label{hdrician},
\end{align}
\end{subequations}
where $\rho$ denotes the path loss between the BS and the UE, and $\kappa_k$ and $\xi_k$ denote the path losses between the $k$-th RIS and the UE/BS. 
Here, $\bm{\tilde{h}}^{\textrm{NLOS}}_{{\rm d}}$,
$\bm{\tilde{h}^{\textrm{NLOS}}}_{{\rm r},k}$ and $\bm{\tilde{G}}_{{\rm r},k}^{\textrm{NLOS}}$ denote the non-line-of-sight (NLoS) components and their entries are generated independently according to $\mathcal{C}\mathcal{N}(0,1)$. 

The line-of-sight (LoS) component of the reflection channel contains angular information about the location of UE. 
For example, $\bm{\tilde{h}}_{{\rm r},k}^{\rm LOS}$ is a function of UE location and $k$-th RIS location. Let $\phi^{\rm RIS}_k$ and $\psi^{\rm RIS}_k$ denote the azimuth and elevation angles of arrival from the UE to the $k$-th RIS. 
The LoS component of $\bm{{h}}_{{\rm r},k}$ can be written as
\begin{equation}\label{ris_to_ue}
    \bm{\tilde h}_{{\rm r},k}^{\rm LOS} = \bm{a}^{\rm RIS}(\phi_k^{\rm RIS}, \psi_k^{\rm RIS}),
\end{equation}
\noindent where the steering vector of the $n$-th element of the $k$-th RIS can be expressed as \cite{taojournal} \cite{multiAIRS}
\begin{align*}
[ \bm{a}^{\rm RIS} & (\phi_k^{\rm RIS}, \psi_k^{\rm RIS})]_n = \\
&
e^{\scriptsize j\Lambda \{ v_1(n, C_k) {\rm sin}(\phi_k^{\rm RIS}){\rm cos}(\psi_k^{\rm RIS}) + v_2(n, C_k) {\rm sin}(\psi_k^{\rm RIS}) \}}, \label{eqn} \numberthis 
\end{align*}
where $\Lambda =  \begin{array}{ll} {2\pi d_{\rm R}} / {\lambda_c}  \end{array}$.
Here, $d_{\rm R}$ is the distance between two reflective elements of the RIS, $\lambda_c$ is the carrier wavelength, and $v_1(n,C_k) = {\rm mod}(n-1,C_k)$ and $v_2(n, C_k) = \lfloor{\frac{n-1}{C_k}}\rfloor$. We use $C_k$ to denote the number of columns of the $k$-th RIS.

Let $\eta_k^{\rm RIS}$ and $\vartheta_k^{\rm RIS}$ denote the azimuth and elevation AoD from the $k$-th RIS to the BS, and let $\psi_k^{\rm BS}$ denote the elevation AoA from the $k$-th RIS to the BS.
The LoS component of $\bm{G}_{{\rm r},k}$ is given by 
\begin{equation}
    \bm{\tilde G}_{{\rm r},k}^{\rm LOS} = ~ \bm{a}^{\rm BS}( \psi_k^{\rm BS}) \bm{a}^{\rm RIS}(\eta_k^{\rm RIS}, \vartheta_k^{\rm RIS})^{\rm H},
\end{equation}
where the steering vector of the BS is given by
\begin{equation}\label{bssteeringvec}
\bm{a}^{\rm BS} ( \psi_k^{\rm BS}) = \left[1 , \cdots, e^{\scriptsize \begin{array}{ll} j\dfrac{2\pi (M-1)d_{\rm A}}{\lambda_c} \cos(\psi_k^{\rm BS}) \end{array}}  \right]^\top,
\end{equation}
where $d_{\rm A}$ denotes the distance between two adjacent antenna elements. 
Lastly, let $ \psi^{\rm UE}$ denote the elevation AoA from the UE to the BS, the LoS component of $\bm{h}_{{\rm d}}$ is given by 
\begin{equation}
\bm{\tilde{h}}_{{\rm d}}^{\rm LOS} = \bm{a}^{\rm BS} ( \psi^{\rm UE}).
\end{equation}

\subsection{Signal Transmission and Reception}

Upon a localization request, the UE 
sends a sequence of $T$ uplink pilot symbols to the BS over $T$ time frames. 
Let $x^{(t)}\in\mathbb{C}$ be the pilot symbol to be transmitted from the UE to the BS at the $t$-th time frame. Due to the limited RF chain constraint, the BS must use an analog sensing vector ${\bm w}^{(t)}$ to sense the channel. The received signal is a combination of the signal from the direct path and the signal reflected off the RISs, so the received pilot at the BS can be expressed as
\begin{equation}\label{ris_pilot}
\begin{split}
    y^{(t)}(\bm{w}^{(t)},& 
	[\bm{\theta}^{(t)}_{k}]_{k=1}^{K})=  \\ & 
\sqrt{P_u}(\bm{w}^{(t)})^\top(\bm{h}_{\rm d} + \sum_{k=1}^K\bm{H}_{{\rm c},k}^\top \bm{\theta}^{(t)}_{k})x^{(t)}+ n^{(t)}, 
\end{split}
\end{equation}
where $P_u$ is the uplink transmission power, $\bm{H}_{{\rm c},k} = \text{diag}(\bm{h}_{{\rm{r}},k})\bm{G}_{{\rm{r}},k}^\top\in\mathbb{C}^{N_k\times M}$ is the cascade channel between the BS and the UE through the reflection at the $k$-th RIS, and $n^{(t)} \sim\mathcal{C}\mathcal{N}(0,\sigma_u^2)$ is the uplink additive white Gaussian noise. 
Here, the received pilot is a function of the analog beamforming vector $\bm{w}^{(t)}$ and the set of $K$ RIS configurations at the $t$-th time frame $[\bm{\theta}^{(t)}_{k}]_{k=1}^{K}$, with the constraint
\begin{equation}
    \left|[\bm{\theta}^{(t)}_{k}]_n\right| = 1, \;\; \; \text{and} \;\; \;  \left|[\bm{w}^{(t)}]_m\right| = 1.
\end{equation}
The main objective of this paper is to design the analog beamforming vector and reflection coefficients actively to enable better localization accuracy (instead of configuring the reflection coefficients randomly as in most of the conventional methods).



\subsection{Localization with RIS}

The goal of the localization problem is to estimate the unknown UE position $\bm{p}$ based on the $T$ observations $[{y}^{(t)}]_{t=0}^{T-1}$, and the known BS and RISs positions. 
The RIS is a cost-effective way of collecting directional information 
due to the fact that it has a large number of elements. The RIS can in fact be seen as a way to deploy additional effective anchor points. 
Although the RIS cannot directly transmit or receive signals, by reconfiguring 
its reflection coefficients over multiple time slots and by analyzing the
received signals at the BS, it can significantly aid the localization task.
For example, it has been advocated in \cite{nguyen,Ubiquitous,Ubiquitous2} that by strategically selecting a fixed set of RIS configurations, a more favourable RSS distribution can be obtained, which allows for improvement in localization accuracy. Here, favourable means that the RSS values at adjacent sampling locations become more diverse and easier to distinguish. 

However, most existing works are based on the \textit{fixed} sensing framework, which passively collects all the observations of the environment according to a fixed set of RIS configurations.
 In essence, the RIS probes the search area using fixed beams along multiple random directions.
 In this paper, we instead propose an \textit{active} sensing framework to gradually narrow down the searching area using more directional beams. This is achieved by designing the RIS reflection coefficients and beamforming vector sequentially based on received pilots so far. 



Specifically, we consider the following active localization setup. 
In the $t$-th time frame, the BS designs the next set of $K$ RIS configurations $[\bm{\theta}^{(t+1)}_{k}]_{k=1}^{K}$ and the next beamforming vector $\bm{w}^{(t+1)}$ based on the existing observations, i.e., $[{y}^{(\tau)}]_{\tau=0}^{t}$. 
Thus, the design of sensing vectors (including the RIS reflection coefficients and the BS beamforming vector) for the $(t+1)$-th stage 
is a function of historical measurements:
\begin{equation}\label{function_g}
   \left[ \bm{w}^{(t+1)},[\bm{\theta}^{(t+1)}_{k}]_{k=1}^{K}  \right]= \mathcal{G}^{(t)}\left([{y}^{(\tau)}]_{\tau=0}^{t} \right),
\end{equation}
where $\mathcal{G}^{(t)}:\mathbb{C}^{t+1} \rightarrow \mathbb{C}^{M}\times \mathbb{C}^{N_1} \times \cdots \times \mathbb{C}^{N_K}$ denote the mapping from existing received pilots to the next sensing vectors. The designed sensing vector from (\ref{function_g}) is subsequently used to make the next measurement ${y}^{(t+1)}$ in the $(t+1)$-th time frame. 
As no prior observation exists when $t < 0$, the first set of beamforming vector $\bm{w}^{(0)}$ and RIS configurations $\bm{\theta}_k^{(0)}$ are produced via function $\mathcal{G}^{(-1)}(\emptyset)$. The function accepts an empty set as input and always produces the same initialization of $K$ RIS configurations and the beamforming vector at the BS.


After receiving all the pilots, the estimated UE position $\hat{\bm{p}}$ is produced as a function of all $T$ historical observations: 
\begin{equation} \label{function_f}
    \hat{\bm{p}}  = \mathcal{F}\left([{y}^{(t)}]_{t=0}^{T-1}\right),
\end{equation}
where $\mathcal{F}:\mathbb{C}^{T} \rightarrow  [\hat{x},\hat{y},\hat{z}]^\top$ denotes the mapping from all received pilots to estimated UE 3D position. 
The problem of minimizing localization error can now be formulated as:
\begin{subequations} \label{prob_formulation}
\begin{align}
\hspace*{-1em}
\underset{\scriptsize \begin{array}{ll} \{{\mathcal{G}}^{(t)}(\cdot)\}_{t=0}^{T-1}, \mathcal{F}(\cdot)  \end{array} }{\textrm{minimize}}&  \mathbb{E}\left[ \| \hat{\bm{p}}- \bm{p} \|_2^2 \right]\label{prob_formulation_obj}\\
\textrm{subject to} \quad \:\:\: & \left|[\bm{\theta}^{(t+1)}_{k}]_n\right|= 1, ~\forall n,k,t,\\
&\left|[\bm{w}^{(t+1)}]_m\right|= 1, ~\forall m, t, \\
& \left[  \bm{w}^{(t+1)}, [\bm{\theta}^{(t+1)}_{k}]_{k=1}^{K} \right]= \mathcal{G}^{(t)}\left([{y}^{(\tau)}]_{\tau=0}^{t}\right),\label{prob_formulation_g}\\
& \hat{\bm{p}}  = \mathcal{F}\left([{y}^{(t)}]_{t=0}^{T-1}\right).
\end{align}
\end{subequations}

Solving the optimization problem (\ref{prob_formulation}) is challenging, as
the problem amounts to optimizing the functions 
$\{\mathcal{G}^{(t)}(\cdot)\}_{t=0}^{T-1}$ in (\ref{function_g}) and $\mathcal{F}(\cdot)$ in
(\ref{function_f}). To make the problem more tractable, a common approach is to
select the beamforming vector and the set of RIS configurations adaptively from a predefined \emph{codebook}, based on heuristics. For example, \cite{hejiguang}
considers an RIS-assisted localization problem in a 2D setting, where the sequence of RIS
patterns is heuristically selected from a hierarchical codebook, based on measurements made so far. 
However, the codebook-based approach is not ideal as the
freedom of designing RIS configurations is restricted. 

In this paper, we pursue a codebook-free approach to solving the above problem.
To this end, one strategy is to tackle the problem in a greedy fashion, i.e.,
to minimize the MSE in each stage. Because computing the exact MSE
objective (\ref{prob_formulation_obj}) involves high dimensional integration
which is not easy to perform numerically, prior works often resort to using BCRLB 
as a metric to minimize at each sensing stage.  This strategy has been proposed in \cite{adaptiveMIMO} and is discussed in detail in Section \ref{sec.crlb}.  
It is used as a comparison point for the proposed data-driven strategy.
In the proposed strategy, we directly solve (\ref{prob_formulation}) and use a neural network to parameterize the function $\mathcal{F}(\cdot)$ and $\mathcal{G}^{(t)}(\cdot)$ in (\ref{prob_formulation}) to adaptively design a sequence of sensing vector.
This learning-based framework can significantly
outperform the BCRLB-based design as illustrated by the simulation results in Section~\ref{sec.numerical_irs_single}. 

\begin{figure*}[!t]
    \centering
  \includegraphics[width=\textwidth]{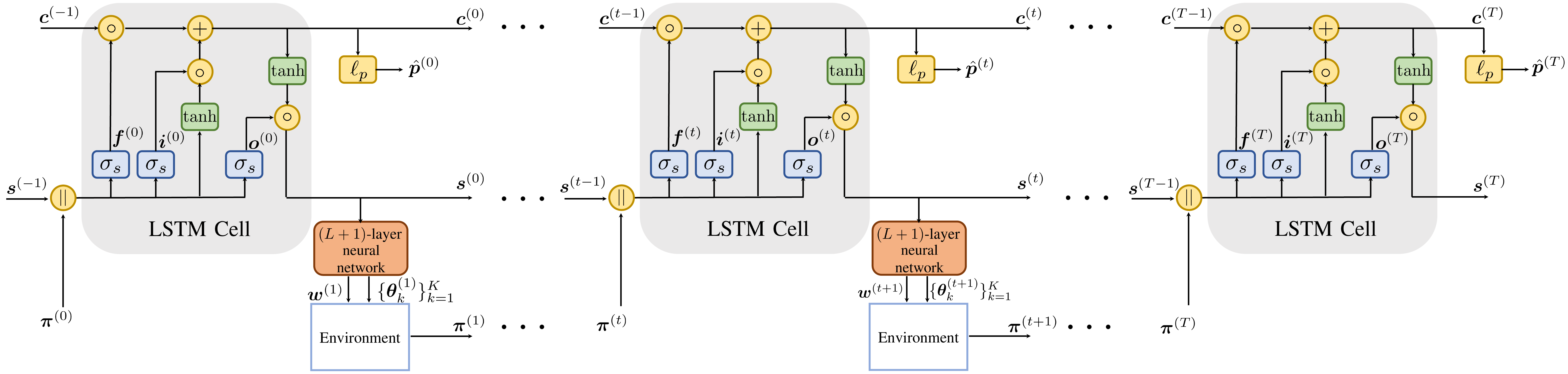}%
  \caption{Proposed active localization framework.}
  \label{fig.lstmcell}%
  \vspace{-0.3em}
\end{figure*}

\section{Proposed Data-Driven Active Localization Framework}
\label{sec.rnn}


In this section, we propose a deep active sensing approach to solve the localization problem discussed in (\ref{prob_formulation}), without assuming a specific geometric channel model and/or restricting the design space of RIS coefficients to a pre-defined codebook. 
Here, as in problem (\ref{prob_formulation}), the next beamforming vector $\bm{w}^{(t+1)}$ and the next set of $K$ RIS configurations $[\bm{\theta}_k^{(t+1)}]_{k=1}^K$ are designed based on historical measurements up to $t$, i.e., 
$ [{y}^{(\tau)}]_{\tau=0}^{t}$. 
The main design challenge is that as the dimension of historical observations increases linearly with $t$, it is difficult to use the entire history of measurements to design the next beamforming vector and the next set of $K$ RIS configurations for large $t$. 
Thus, the key design question for the neural network is how to extract
useful information from the historical measurements and how to map it to 
an information vector of fixed dimension.

In this paper, we develop an LSTM network capable of automatically constructing an information vector of fixed dimension from existing measurements and extracting temporal features and long-term dependencies from a sequence of temporal inputs. 
The specific LSTM network developed here is similar to the one in
\cite{sohrabi2021active} in which the information vector is the hidden state,
but with modifications to the neural network architecture tailored to the localization problem in an RIS-assisted network. 
We also propose new loss functions to make
the LSTM network generalizable to the number of time frames.

\subsection{Neural Network Architecture}

The overall neural network architecture is shown in Fig.~\ref{fig.lstmcell}. 
At the $t$-th time frame, an LSTM cell takes new features as input to update the hidden state vector $\bm{s}^{(t)}$ and the cell state vector $\bm{c}^{(t)}$. The new input, denoted as $\bm{\pi}^{(t)}$, can be the RSS $|{y}^{(t)}|^2$
or the real and imaginary component of received pilots $[\mathcal{R}({y}^{(t)}), \mathcal{I}({y}^{(t)})]$, depending on the hardware constraints at the receiver.
The updating rules of the cell state vector $\bm{c}^{(t)}$ and hidden state vector $\bm{s}^{(t)}$ are as follows:
\begin{subequations} 
\begin{align}
\bm{c}^{(t)} = \; & \bm{f}^{(t)} \circ \bm{c}^{(t-1)} + \bm{i}^{(t)} \circ {\rm tanh}\left({\bm{{u}}}_c(\bm{\pi}^{(t)}) + {\bm{{r}}}_c(\bm{s}^{(t-1)})\right),\\
\bm{s}^{(t)} = \; & \bm{o}^{(t)} \circ {\rm tanh}(\bm{c}^{(t)}),
\end{align}
\end{subequations}
where ${\bm{{u}}}_c(\cdot)$ and ${\bm{{r}}}_c(\cdot)$ are linear layers with a number of fully connected units. Here, $\bm{f}^{(t)}$, $\bm{i}^{(t)}$, and $\bm{o}^{(t)}$ are the activation vectors of the forget gate, input gate and output gate within the LSTM cell respectively. The element-wise updating rules of the different gates are as follows:
\begin{subequations}\label{fio_function}
\begin{align}
\bm{f}^{(t)} =\;& \sigma_s\left({\bm{{u}}}_f(\bm{\pi}^{(t)})+{\bm{{r}}}_f(\bm{s}^{(t-1)})\right),\\
\bm{i}^{(t)} =\;& \sigma_s\left({\bm{{u}}}_i(\bm{\pi}^{(t)})+{\bm{{r}}}_i(\bm{s}^{(t-1)})\right),\\
\bm{o}^{(t)} =\;& \sigma_s\left({\bm{{u}}}_o(\bm{\pi}^{(t)})+{\bm{{r}}}_o(\bm{s}^{(t-1)})\right),
\end{align}
\end{subequations}
\noindent where $\sigma_s(\bm{x}) = [1/(1+e^{-\bm{x}_1}), \cdots, 1/(1+e^{-\bm{x}_i}),\cdots]^\top$ is the element-wise sigmoid function, and ${\bm{{u}}}_f$, ${\bm{{u}}}_i$, ${\bm{{u}}}_o$, ${\bm{{r}}}_f$, ${\bm{{r}}}_s$ and ${\bm{{r}}}_o$ are linear layers. 
In Fig.~\ref{fig.lstmcell}, we use $||$ to denote the update-then-sum operations on $\bm{\pi}^{(t)}$ and $\bm{s}^{(t-1)}$ within the element-wise sigmoid function, i.e., ${\bm{{u}}}_{\varrho}(\bm{\pi}^{(t)})+{\bm{{r}}}_{\varrho}(\bm{s}^{(t-1)})$, where $ \varrho \in \{ f,i,o\}$ in (\ref{fio_function}). 
As per convention, the initial value of the cell state $\bm{c}^{(0)}$ is obtained by setting $\bm{s}^{(-1)} = \bm{c}^{(-1)}=\bm{0}$, and ${y}^{(0)} = 1$. 


We design the sensing vector and RIS configurations for time frame $t+1$ based on the hidden state vector $\bm{s}^{(t)}$. The hidden state vector is used as the input to a fully connected neural network of $L$ layers to obtain the right representation of information to design the beamforming vector and the set of RIS configurations for the next time frame
\begin{equation}
    \bm{\gamma}^{(t)} = \beta_{L}(\bm{A}_L \beta_{L-1}(\cdots \beta_1(\bm{A}_1\bm{s}^{(t)}+\bm{b}_1)\cdots)+\bm{b}_L),
\end{equation}
\noindent where $\beta_l$, $l\in \{1,\cdots,L\}$ is the activation function of the $l$-th layer, which is set to ${\rm relu}(x) = {\rm max}(0,x)$, $\{\bm{A}_l\}_{l=1}^{L}$ and $\{\bm{b}_l\}_{l=1}^{L}$ are sets of trainable weights and biases. Subsequently, we obtain the RIS configuration and the beamforming vector from $\bm{\gamma}^{(t)}$ using the $L+1$-th layer
\begin{subequations} 
\begin{align}
\bm{\bar{\theta}}^{(t+1)} = \; & \bm{\dot{A}}_{L+1}\bm{\gamma}^{(t)}+ \bm{\dot{b}}_{L+1},\\
\bm{\bar w}^{(t+1)} = \; & \bm{\ddot{A}}_{L+1}\bm{\gamma}^{(t)}+ \bm{\ddot{b}}_{L+1}.
\end{align}
\end{subequations}
Here, $\bm{\bar{\theta}}^{(t+1)}$ contains sequentially ordered real and imaginary components of the set of RISs configurations, and $\bm{\bar w}^{(t+1)}$ contains the real and imaginary components of the beamforming vector
\begin{subequations} 
\begin{align}
\bm{\bar{\theta}}^{(t+1)} = \; & [ \mathcal{R}(\bm{{\theta}}^{(t+1)}_{1})^\top, \mathcal{I}(\bm{{\theta}}^{(t+1)}_{1})^\top,\cdots,
 \mathcal{I}(\bm{{\theta}}^{(t+1)}_{K})^\top]^\top,\\
\bm{\bar w}^{(t+1)} = \; & [ \mathcal{R}(\bm{w}^{(t+1)})^\top,
 \mathcal{I}(\bm{w}^{(t+1)})^\top]^\top.
\end{align}
\end{subequations}
The dimensions of trainable weights and biases, $\bm{\dot{A}}_{L+1}$, $\bm{\ddot{A}}_{L+1}$, $\bm{\dot{b}}_{L+1}$ and $\bm{\ddot{b}}_{L+1}$, are designed to ensure the output is of correct dimension, such that $\bm{\bar w}^{(t+1)} \in\mathbb{R}^{2M}$ and $\bm{\bar{\theta}}^{(t+1)} \in\mathbb{R}^{2\Delta}$, where $\Delta= \sum_k N_k$. 
To enforce unit modulus constraint on each element of the sensing vector the RIS reflection coefficients, an element-wise normalization is performed
\begin{subequations} 
\begin{align}
\begin{split}
    [\bm{\theta}^{(t+1)}_{k}]_n = \: & \dfrac{[\mathcal{R}(\bm{{\theta}}^{(t+1)}_{k})]_{n}}{\sqrt{[\mathcal{R}(\bm{{\theta}}^{(t+1)}_{k})]_{n}^2+[\mathcal{I}(\bm{{\theta}}^{(t+1)}_{k})]_{n}^2}} \\ & +  j\dfrac{[\mathcal{I}(\bm{{\theta}}^{(t+1)}_{k})]_{n}}{\sqrt{[\mathcal{R}(\bm{{\theta}}^{(t+1)}_{k})]_{n}^2+[\mathcal{I}(\bm{{\theta}}^{(t+1)}_{k})]_{n}^2}},
\end{split}\\
\begin{split}
    [\bm{w}^{(t+1)}]_m =\: &  \dfrac{[\mathcal{R}(\bm{w}^{(t+1)})]_{m}}{\sqrt{[\mathcal{R}(\bm{w}^{(t+1)})]_{m}^2+[\mathcal{I}(\bm{w}^{(t+1)})]_{m}^2}} 
    \\ & +  j\dfrac{[\mathcal{I}(\bm{w}^{(t+1)})]_{m}}{\sqrt{[\mathcal{R}(\bm{w}^{(t+1)})]_{m}^2+[\mathcal{I}(\bm{w}^{(t+1)})]_{m}^2}}.
\end{split}
\end{align}
\end{subequations}






While the hidden state vector $\bm{s}^{(t)}$ is used to design the sensing vectors for the next time frame, the cell state vector $\bm{c}^{(T)}$ is used to obtain the estimated UE position at the $T$-th time frame.
After $T$ time frames, the final estimated UE position $\hat{\bm{p}}^{(T)}$ is obtained through a fully connected neural network, 
based on the final cell state $\bm{c}^{(T)}$:
\begin{equation}
    \hat{\bm{p}}^{(T)} = \ell_p(\bm{c}^{(T)}),
\end{equation}
where $\ell_p(\cdot)$ denote a fully connected neural network. 

\subsection{Loss Functions}

To train the LSTM network, we employ Adam optimizer \cite{adam} to minimize the average mean squared error (MSE) between the estimated position $\hat{\bm{p}}^{(T)}$ and the true position as follows
\begin{equation}\label{pt_mse}
    \begin{split}
    \mathbb{E}\left[ \| \hat{\bm{p}}^{(T)}- \bm{p} \|_2^2 \right].
    \end{split}
\end{equation}
\noindent Here, the choice of the loss function in (\ref{pt_mse}) encourages the LSTM network to design a series of $T$ beamforming vector and $T$ sets of $K$ RIS configurations to minimize localization error at the final time frame. Once the neural network is trained, the trained network can automatically design the series of sensing vectors based on recurring instantaneous measurements for any UE with an arbitrary position within the service area. 
We note that this loss function only accounts for the estimation error at the final stage. This is a good choice since it gives the neural network the freedom to design the sensing strategy across the entire $T$ measurement stages.


In some cases, we may need to have earlier stopping in the sequence of sensing stages, e.g., due to some latency constraints. This requires the neural network to output the best estimates of the location before the final stage, but this is not considered in the loss function \eqref{pt_mse}. For this scenario, we propose a weighted MSE loss function as follows:
\begin{equation}\label{weighted_mse}
    \begin{split}
    \mathbb{E}\left[\sum_{t = 1}^{T} \alpha_t \| \hat{\bm{p}}^{(t)}- \bm{p} \|_2^2 \right],
    \end{split}
\end{equation}
where $\hat{\bm{p}}^{(t)} = \ell_p(\bm{c}^{(t)})$ and the weight $\sum_{t=1}^{T}\alpha_t = 1$. 
By shifting weights to shorter sensing stages where early stopping is expected, we can encourage the LSTM network to design sensing configurations to reduce estimation error at earlier time frames. 
The weighted loss function (\ref{weighted_mse}) allows more flexibility and ensures generalizability by putting non-zero weights in earlier frames.



\section{BCRLB-Based Greedy Active Sensing}
\label{sec.crlb}





A notable feature of the proposed learning-based approach is that it solves
problem (\ref{prob_formulation}) by designing the functions
$\{\mathcal{G}^{(t)}(\cdot)\}_{t=0}^{T-1}$ in (\ref{function_g}) \emph{jointly} across $t$
to minimize the localization error.  If we were to do this analytically, the computational
complexity of such an optimization would grow dramatically as a function of the number of sensing stages.
To keep the problem tractable, an analytical approach would typically need to 
design individual $\mathcal{G}^{(t)}(\cdot)$ for each time frame in a greedy fashion. 
Further, instead of minimizing MSE, which is difficult to compute, an analytic method would typically optimize a lower bound on MSE. 

Nevertheless, such an analytic approach is an important benchmark. 
In this section, we present a model-based analytic approach for solving the active sensing problem 
for localization based on estimation theory. The approach is similar to the one proposed in 
\cite{adaptiveMIMO}. 
In essence, instead of optimizing a direct mapping from the observations to the sensing vector, 
the function expression $\mathcal{G}^{(t)}(\cdot)$ in (\ref{prob_formulation_g}) is cast as 
a two-step mapping from existing observations to the BCRLB, then from the BCRLB to the sensing vector. 
Specifically, 
at each sensing stage, the posterior distribution of the unknown UE position is updated based on existing pilots. The conditional BCRLB is subsequently updated based on the posterior distribution. 
Finally, we optimize the RIS reflection 
coefficients with the objective of minimizing the conditional BCRLB. 

The derivation of the Bayesian Fisher information matrix and the optimization
procedure are presented below for the single-RIS SISO setting as in
\cite{adaptiveMIMO}. This derivation is different from the non-Bayesian CRLB formulation of \cite{liu2020reconfigurable}, which does not account for the
prior distribution of the unknown location.
In this section, we derive the BCRLB for a single-RIS SISO network for simplicity. We note that the conventional CRLB for a single-RIS MIMO network and multi-RIS MIMO system are derived in \cite{liu2020reconfigurable} and \cite{multiriscrlb} respectively.


\subsection{BCRLB Formulation}

Recall that the unknown positional parameters to be estimated are as follows:
\begin{equation}\label{crlb_AoAs}
    \bm{p} = [x,y,z]^\top \in\mathbb{R}^{3} .
\end{equation}
Based on the measurement $[{y}^{(\tau)}]_{\tau=0}^{t-1}$, we are interested in designing the next RIS configuration $\bm{\theta}^{(t+1)}$ to improve the estimation performance of $\bm{p}$ in terms of conditional MSE. In this model-based approach, we solve the active sensing problem (\ref{prob_formulation}) on a per-stage basis
as follows
\begin{subequations} \label{prob_formulation_mse_cond}
\begin{align}
\underset{\scriptsize \begin{array}{ll} {\mathcal{G}}^{(t)}(\cdot), \mathcal{F}(\cdot)  \end{array} }{\textrm{minimize}}&  \underbrace{\mathbb{E}\left[\|\hat{\bm{p}}^{(t)} - \bm{p}\|_2^2 \Big |  [{y}^{(\tau)}]_{\tau=0}^{t} \right]}_{\bm{\Gamma}^{(t)}} \\
\textrm{subject to} \;\;  & \left|[\bm{\theta}^{(t+1)}]_n\right|= 1, ~\forall n,\\
& \bm{\theta}^{(t+1)}= \mathcal{G}^{(t)}\left([{y}^{(\tau)}]_{\tau=0}^{t}\right),\\
& \hat{\bm{p}}^{(t)}  = \mathcal{F}\left([{y}^{(\tau)}]_{\tau =0}^{t}\right),\label{prob_formulation_mse_cond_f}
\end{align}
\end{subequations}
where $\hat{\bm{p}}^{(t)}$ is the estimated positional parameters at the $t$-th time frame. 


While both problem formulations (\ref{prob_formulation}) and  (\ref{prob_formulation_mse_cond}) satisfy the broad definition of active sensing, 
we point out that the design objective in (\ref{prob_formulation_mse_cond}) is for one single sensing stage, i.e., it greedily treats the active sensing problem by designing ${\mathcal{G}}^{(t)}(\cdot)$ at each time frame to minimize the MSE in the immediate next stage, 
instead of designing a joint active sensing strategy across $T$ stages, i.e., $\{{\mathcal{G}}^{(t)}(\cdot)\}_{t=0}^{T-1}$ as in (\ref{prob_formulation}). 
The solution space of the objective (\ref{prob_formulation_mse_cond}) is a subset of the solution space of the objective (\ref{prob_formulation}).

Still, the evaluation of $\bm{\Gamma}^{(t)}$ is not trivial, because high-dimensional integration is involved. 
To address this issue, we consider the conditional BCRLB as an optimization criterion for designing the next sensing vector. The BCRLB provides a lower bound of the MSE that does not depend on the exact value of unknown parameter $\bm{p}$. The conditional BCRLB allows the utilization of posterior information based on the existing measurements \cite{conditional_bcrlb}. In the $t$-th time frame, we let $\bm{{\rm J}}^{(t)}$ denote the conditional Fisher information matrix, and $\bm{\Gamma}^{(t)}$ denote the conditional MSE matrix. The MSE is bounded by the conditional BCRLB 
as follows
\begin{equation}\label{BCRLB}
    \tr\left(\bm{\Gamma}^{(t)}\right) > \tr\left(\left[\bm{{\rm J}}^{(t)}\right]^{-1}\right), 
\end{equation}
where the entries of $\bm{{\rm J}}^{(t)}$ are given by
\begin{equation}\label{crlb_t}
    [\bm{{\rm J}}^{(t)}]_{i,j} =- \mathbb{E}\left[ \dfrac{\partial^2 \log f\left(y^{(t)},\bm{p}|[{y}^{(\tau)}]_{\tau=0}^{t-1}\right)}{\partial {p}_i \partial {p}_j }  
    \right].
\end{equation}
Here, $f(y^{(t)},\bm{p}\big |[{y}^{(\tau)}]_{\tau=0}^{t-1})$ denotes conditional distribution of $y^{(t)}$ and $\bm{p}$ given existing observations $[{y}^{(\tau)}]_{\tau=0}^{t-1}$. The expectation is taken with respect to the conditional distribution $f\left (y^{(t)},\bm{p}\big|[{y}^{(\tau)}]_{\tau=0}^{t-1} \right)$. The terms $p_i$ and $p_j$ refer to any arbitrary two position parameters in $\{x,y,z\}$.

The problem formulation for the BCRLB-based active sensing approach in minimizing the MSE is now given by
\begin{subequations} \label{prob_formulation_bcrlb}
\begin{align}
\hspace*{-1em}
\underset{\scriptsize \begin{array}{ll} {\mathcal{G}}^{(t)}(\cdot) \end{array} }{\textrm{minimize}} \quad &  \tr\left(\left[\bm{{\rm J}}^{(t)}\right]^{-1}\right)\\
\textrm{subject to}  \quad & \left|[\bm{\theta}^{(t+1)}]_n\right|= 1, ~\forall n,\\ 
& \bm{\theta}^{(t+1)}= \mathcal{G}^{(t)}\left([{y}^{(\tau)}]_{\tau=0}^{t}\right).
\end{align}
\end{subequations}
The problem (\ref{prob_formulation_mse_cond}) is solved successively by solving the problem (\ref{prob_formulation_bcrlb}) to design the active sensing vector over $T$ stages. 
After the $T$ sensing stages, the functional mapping ${\mathcal{F}}(\cdot)$ in (\ref{prob_formulation_mse_cond_f}) is designed by adopting a maximum a posteriori estimator to obtain the estimated UE position $\hat{\bm{p}}^{(T)}$. 


\subsection{BCRLB Calculation}

We can make use of Bayes's theorem and decompose (\ref{crlb_t}) as in \cite{adaptiveMIMO}
\begin{equation}
     [\bm{{\rm J}}^{(t)}]_{i,j} = [\bm{{\rm J}}_{\rm D}^{(t)}]_{i,j} + [\bm{{\rm J}}_{\rm P}^{(t-1)}]_{i,j},
\end{equation}
where $\bm{{\rm J}}_{\rm D}^{(t)}$ denote the incremental Bayesian Fisher information matrix based on the new observation,
and $\bm{{\rm J}}_{\rm P}^{(t-1)}$ is the Fisher information matrix based on historical observations.
To write down $[\bm{{\rm J}}_{\rm D}^{(t)}]_{i,j}$ and $[\bm{{\rm J}}_{\rm P}^{(t-1)}]_{i,j}
$ explicitly, 
recall from (\ref{ris_pilot}) that $\bm{H}_{{\rm c}, k} \in \mathbb{C}^{N_k \times M}$ denotes the MISO cascade channel between the BS and the UE through the reflection at the $k$-th RIS. Here, we remove the subscript $k$ as there is only one RIS considered here and let $\bm{h}_{{\rm c}} \in \mathbb{C}^{N}$ denote the SISO cascade channel through the reflection at the RIS. The entries of the incremental Bayesian Fisher information matrix are given by
\begin{subequations}\label{J_D}
\begin{align}
[\bm{{\rm J}}_{\rm D}^{(t)}]_{i,j} = ~ & - \mathbb{E}\left[ \dfrac{\partial^2 \log f(y^{(t)}\big|\bm{p},[{y}^{(\tau)}]_{\tau=0}^{t-1})}{\partial p_i \partial p_j }  
\right]\\
= ~ & \mathbb{E}\left[ - \mathbb{E}\left[ \dfrac{\partial^2 \log f(y^{(t)}\big|\bm{p},[{y}^{(\tau)}]_{\tau=0}^{t-1})}{\partial p_i \partial p_j  } \label{J_D_b} 
\right]
\right]\\
= ~ &  (\bm{\theta}^{(t)})^{\rm H} \mathbb{E}\left[ \dfrac{2}{\sigma_u^2}  \left[ \bm{h}_{\rm c}^\prime(p_i) (\bm{h}_{\rm c}^\prime(p_j))^{\rm H}  \right] 
\right]\bm{\theta}^{(t)},
\end{align}
\end{subequations}
where 
$\bm{h}_{\rm c}^\prime(p_i) = {\partial \bm{h}_{\rm c}(p_i)}/{\partial p_i}$. 
Here in (\ref{J_D_b}), the expectations are taken with respect to 
$f\left (\bm{p}\big|[{y}^{(\tau)}]_{\tau=0}^{t-1} \right)$ and $f\left (y^{(t)} \big|\bm{p},[{y}^{(\tau)}]_{\tau=0}^{t-1} \right)$ respectively. Note that we have only accounted for the cascade channel and not the direct channel between the BS and the UE because the BS and UE are equipped with a single antenna, hence it is not necessary to account for the elevation AoA from the UE to the BS, i.e., $\psi^{\rm UE}$, as the LoS component of $h_d$ reduces to 
${\tilde{h}}_{{\rm d}}^{\rm LOS} = \bm{a}^{\rm BS} ( \psi^{\rm UE}) = 1$. 
The entries of the Fisher information matrix based on the historical observations are given by
\begin{equation}\label{J_P_entries}
\begin{split}
     [\bm{{\rm J}}_{\rm P}^{(t-1)}]_{i,j} = & - \mathbb{E}\left[ \dfrac{\partial^2 \log f\left(\bm{p}|[{y}^{(\tau)}]_{\tau=0}^{t-1}\right)}{\partial {p}_i \partial {p}_j } \right]\\
    = & \sum_{t^\prime=1}^{t-1} -\mathbb{E}\left[ \dfrac{\partial^2 \log f\left(y^{(t^\prime)}\big|\bm{p},[{y}^{(\tau)}]_{\tau=0}^{t^\prime-1}\right)}{\partial {p}_i \partial {p}_j }
    \right] \\
    & - \mathbb{E}\left[ \dfrac{\partial^2 \log f\left(\bm{p}\right)}{\partial {p}_i \partial {p}_j } \right],
\end{split}
\end{equation}
where the expectations are taken with respect to $f\left (y^{(t)},\bm{p}\big|[{y}^{(\tau)}]_{\tau=0}^{t-1} \right)$ here. 


\begin{algorithm}[t]
  \caption{BCRLB Localization Algorithm}
    \label{BCRLB_algorithm}
\begin{algorithmic}[1]
    \item Initialize $t=0$; 
    \item \label{algo:posterior} Produce samples from the posterior distribution $f(\bm{p} \big| [{y}^{(\tau)}]_{\tau=0}^{t}, [\bm{\theta}^{(\tau)}]_{\tau =0}^t )$;
    \item Compute entries of $\bm{{\rm J}}_{\rm D}^{(t)}$ in (\ref{J_D}) and $\bm{{\rm J}}_{\rm P}^{(t-1)}$ in (\ref{J_P_entries});
    \item Solve the optimization problem in (\ref{JD_maximize}) to design RIS configuration $\bm{\theta}^{(t+1)}$;
    \item $t$ $\longleftarrow$ $t+1$; repeat Step \ref{algo:posterior} until $t=T$
    \item Obtain estimated position $\hat{\bm{p}}^{(T)}$ from 
    the posterior distribution. 
\end{algorithmic}
\end{algorithm}

It should be noted that the conditional Fisher information matrix in (\ref{BCRLB}) is only a function of $\bm{\theta}^{(t)}$ through $\bm{{\rm J}}_{\rm D}^{(t)}$, as discuss in \cite{adaptiveMIMO}. The optimization problem can be formulated as 
\begin{subequations}\label{JD_maximize}
\begin{align}
\hspace*{-1em}
\underset{\scriptsize \begin{array}{ll}  \bm{\theta}^{(t+1)}  \end{array} }{\textrm{minimize}}  \quad &  \tr\left(\left[\bm{{\rm J}}_{\rm D}^{(t)} +\bm{{\rm J}}_{\rm P}^{(t-1)} \right]^{-1}\right)\\
\textrm{subject to} \quad & \left|[\bm{\theta}^{(t+1)}]_n\right|= 1, ~\forall n.
\end{align}
\end{subequations}
To solve problem (\ref{JD_maximize}), 
we use projected gradient descent as in \cite{hguo}. 
We also need to compute the posterior distribution $f(\bm{p} \big| [{y}^{(\tau)}]_{\tau=0}^{t}, [\bm{\theta}^{(\tau)}]_{\tau =0}^t )$. This can be done recursively
\begin{equation}
\begin{split}
    f(\bm{p} \big| [y^{(\tau)}]_{\tau=1}^t, &[\bm{\theta}^{(\tau)}]_{\tau=1}^t)  \\
    = {C} & \cdot  f( y^{(t)} \big | \bm{p}, \bm{\theta}^{(t)}) f( \bm{p} \big | [{y}^{(\tau)}]_{\tau=0}^{t-1}, [\bm{\theta}^{(\tau)}]_{\tau =0}^{t-1}),
\end{split}
\end{equation}
where ${C} = f([{y}^{(\tau)}]_{\tau=0}^{t-1}, [\bm{\theta}^{(\tau)}]_{\tau =0}^t) / f([{y}^{(\tau)}]_{\tau=0}^{t}, [\bm{\theta}^{(\tau)}]_{\tau =0}^t)$. In this setup, the probability distribution $f( y^{(t)} \big | \bm{p}, [{y}^{(\tau)}]_{\tau=0}^{t-1}, [\bm{\theta}^{(\tau)}]_{\tau =0}^t)$ is complex Gaussian $\mathcal{C}\mathcal{N}( (\bm{h}_{\rm c}(\bm{p}))^{\rm H}\bm{\theta}^{(t)} ,\sigma_u^2)$.
The overall BCRLB-based active sensing algorithm is summarized in Algorithm \ref{BCRLB_algorithm}.



\begin{figure}[t]
    \centering
\includegraphics[width=\columnwidth]{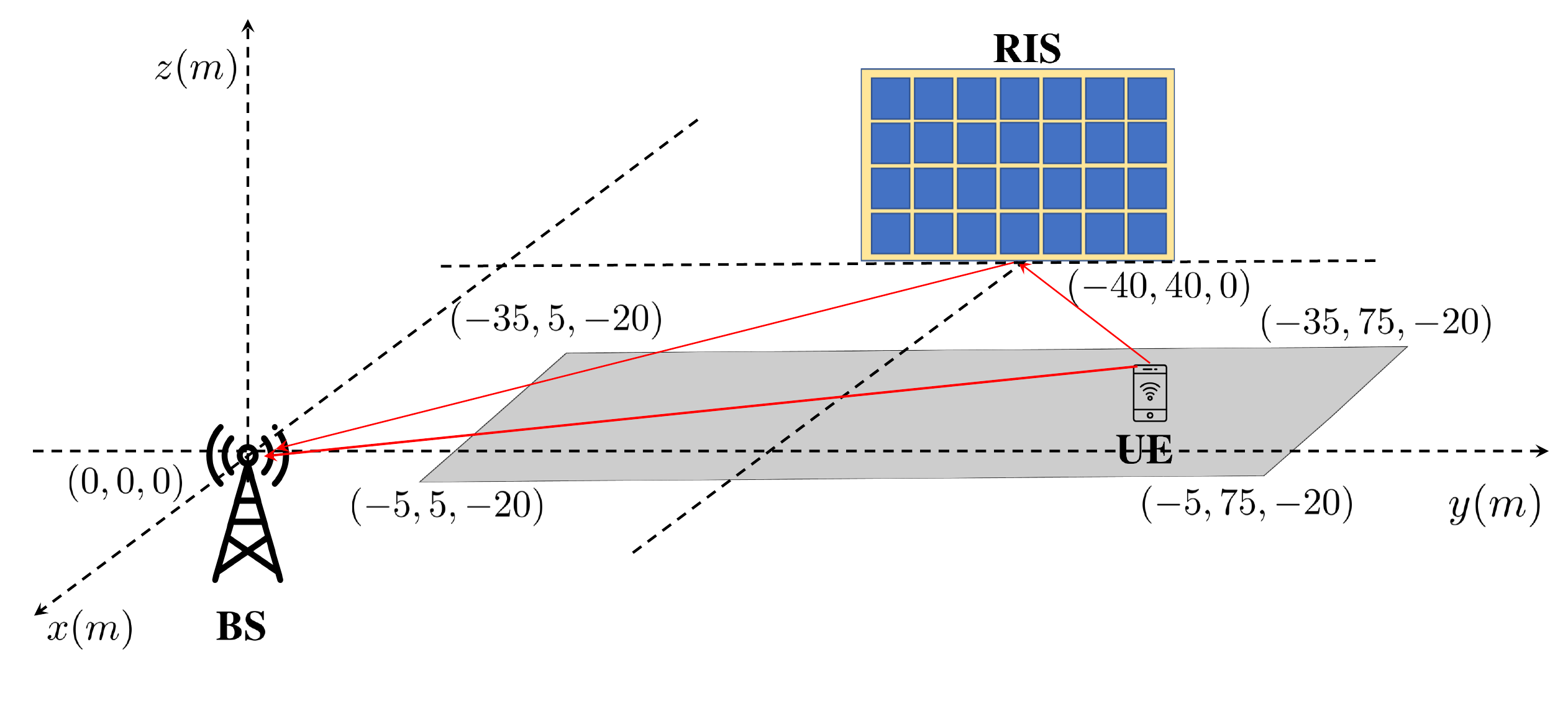}%
\caption{Simulation setting for single-RIS SISO network.}%
\label{fig.SISOsetup_simulation}%
\end{figure}

\section{Numerical Results for Single-RIS SISO Network}
\label{sec.numerical_irs_single}
\subsection{Simulation Environment} \label{siso_singleris_environment}
We first consider the case where only $K=1$ RIS is deployed in the system. The case with multiple RISs is considered in Section~\ref{sec.numerical_irs_multi}. In simulations,
the single-antenna BS is located at $\bm{p}^{\rm BS} = (0m,0m,0m)$. The location of the RIS is $\bm{p}^{\rm RIS} = (-40m,40m,0m)$ as shown in Fig.~\ref{fig.SISOsetup_simulation}. The RIS consists of $8\times 8$ reflective elements. 
The unknown user locations $\bm{p}$ are uniformly generated within a rectangular area on the $x$-$y$ plane $(-20m\pm15m, 40m\pm35m, -20m)$ with $z = -20 m$. 
In subsequent simulation, the Rician factor $\epsilon$ is set to $10$, and $ {2\pi d_{R}}/{\lambda_c} = 1$ without loss of generality. The path loss models of the direct and reflected paths are $32.6+36.7\log(d_1)$ and $30+22\log(d_2)$, respectively, where $d_1$ and $d_2$ denote the corresponding link distance. The bandwidth is $10$ MHz with a background noise of $-170$ dBm/Hz.

Recall that $\phi_k^{\rm RIS}$ and $\psi_k^{\rm RIS}$ denote the azimuth and elevation AoA from the UE to the $k$-th RIS. 
As shown in Fig.~\ref{fig.setup_intro} and Fig.~\ref{fig.SISOsetup_simulation}, the following mapping from $\{\phi_k^{\rm RIS}, \psi_k^{\rm RIS} \}$ to the position of UE can be established:
\begin{subequations}
\begin{align}
\cos(\phi_k^{\rm RIS})\sin(\psi_k^{\rm RIS})=\;& \dfrac{x - x_k^{\rm RIS}}{d_k^{\rm RU}},\\
\sin(\phi_k^{\rm RIS})\sin(\psi_k^{\rm RIS})=\;& \dfrac{y - y_k^{\rm RIS}}{d_k^{\rm RU}},\\
\cos(\psi_k^{\rm RIS})=\;& \dfrac{z_k^{\rm RIS} - z}{d_k^{\rm RU}},
\end{align}
\end{subequations}
where $d_k^{\rm RU}$ denotes the distance between the $k$-th RIS and the UE. Similarly, recall that $\eta_k^{\rm RIS}$ and $\vartheta_k^{\rm RIS}$ denote the azimuth and elevation AoD from the $k$-th RIS to the BS, and let $\psi_k^{\rm BS}$ denote the elevation AoA from the $k$-th RIS to the BS. We can establish the following mapping based on $\{\eta_k^{\rm RIS},\vartheta_k^{\rm RIS},\psi_k^{\rm BS} \}$: 
\begin{subequations}
\begin{align}
\sin(\vartheta_k^{\rm RIS})\cos(\eta_k^{\rm RIS})=\;& \dfrac{x_k^{\rm RIS} - x^{\rm BS}}{d_k^{\rm BR}},\\
\sin(\vartheta_k^{\rm RIS})\sin(\eta_k^{\rm RIS})=\;& \dfrac{y_k^{\rm RIS} - y^{\rm BS}}{d_k^{\rm BR}},\\
\sin(\psi_k^{\rm BS})=\;& \dfrac{z_k^{\rm RIS} - z^{\rm BS}}{d_k^{\rm BR}},
\end{align}
\end{subequations}
where $d_k^{\rm BR}$ denotes the distance between the BS and the $k$-th RIS.

\begin{table}[t]
\vspace{+2em}
\begin{center}
\captionof{table}{Parameters of the LSTM Network.\label{Tab:parameter}}
\begin{tabular}{| c|c |} 
\hline
\textbf{Label}& \textbf{Dimension}\\
\hline
\shortstack{\:\\${\bm{{u}}}_c,{\bm{{u}}}_f,{\bm{{u}}}_i,{\bm{{u}}}_o$ \\ ${\bm{{r}}}_c,{\bm{{r}}}_f,{\bm{{r}}}_i,{\bm{{r}}}_o$ } &\shortstack{ $512$ \\ \quad \quad }\\
\hline
\quad\shortstack{\: \\$\bm{A}_{1}$}$ \quad\quad\;\;\, $\vline $ \quad\quad \{\bm{A}_l\}_{l=2}^{L}   $   & \quad $512 \times 1024 \quad\quad $\vline $ \; \;\;\quad  1024 \times 1024$ \\
\hline
\quad\shortstack{\: \\ $\bm{\dot{A}}_{L+1}$} $\quad $\vline $ \quad\quad\quad $\shortstack{\: \\ $\bm{\ddot{A}}_{L+1}$}   & \quad $1024 \times 2\Delta \quad\quad  $\vline $ \, \quad\quad 1024 \times 2 M$ \\
\hline
\shortstack{\: \\ $\{\bm{b}_l\}_{l=1}^{L}  $}   
\vline  \; \shortstack{ \: \\ $\bm{\dot b}_{L+1}$}  \; 
\vline  \;\;\;\;
\shortstack{\: \\ $\bm{\ddot b}_{L+1}$}   & $ 1024 \quad\quad \; $\vline $\quad \; \; 2\Delta  \;\;\quad $\vline $ \;\quad\quad   2 M$ \\
\hline
$\ell_{p}(\cdot)$   & $ 512 \times 200 \times 200 \times 3$\\
\hline
\end{tabular}
\end{center}
\end{table}

\subsection{Proposed Scheme versus Baseline Schemes}
The proposed LSTM network for adaptive RIS-assisted localization is
implemented using the parameters in Table \ref{Tab:parameter}. We set  $L = 4$ and train the model using Tensorflow \cite{tensorflow}. During the training phase, the model samples a total of 2,048,000 training data in 2000 epochs. The training data include direct channel coefficient and cascade channel vector generated based on arbitrary UE positions; here, the channel coefficients are needed for generating received pilots. 
The training label contains the true UE positions. 
Once the LSTM network is trained, the trained network can automatically design a series of RIS configurations based on recurring instantaneous measurements of the pilots to locate any UE with an arbitrary position within the service area. 
We compare the localization performance of the proposed algorithm against the following baseline schemes. 

\subsubsection{Fingerprint based localization with random sensing configurations\cite{nguyen}}
In this scheme, the sequence of $T$ uplink RIS configurations is fixed to some randomly generated sensing vectors in the sensing phase. This is a nonadaptive localization scheme. Every $1m\times 1 m$ block within the $30m \times 70m$ area is associated with a vector of RSS as its fingerprint, i.e., $\left[|{y}^{(1)}|^2,\cdots,|{y}^{(T)}|^2\right]$.
The fingerprints are collected offline and stored in a database. To perform the localization task, Weighted $k$-nearest neighbour (wKNN) classifier is used to match the received pilots and the true fingerprint stored in the database. Here, we set $k = 5$ in the wKNN algorithm.

\subsubsection{DNN with random or learned RIS configurations}

    The sequence of $T$ uplink RIS configurations is non-adaptive. Here, we can have two different designs for the RIS configurations: \lowerromannumeral{1}) the RIS configurations are randomly chosen, or \lowerromannumeral{2}) the sequence of RIS configurations are learned from training data (i.e., the channel statistics), but are fixed at the testing stage. To realize the scheme \lowerromannumeral{2}) in Tensorflow, we make the reflection coefficients trainable during the neural network training phase so that they can be updated according to the channel distribution to minimize the loss function. A fully connected neural network of dimension $[200,200,200,3]$ is used to map the received pilot symbols over $T$ time frames $[ \mathcal{R}({y}^{(t)}),\mathcal{I}({y}^{(t)})]_{t=0}^{T-1}$ or RSS values over $T$ time frames to an estimated UE position. 

\subsubsection{Optimizing BCRLB using gradient descent (GD)}

Here, we test the idea of designing an active sensing strategy based on minimizing
BCRLB in every time step as discussed in Section \ref{sec.crlb}. 
Here, we discretize the posterior probability distribution of the service area ($30m \times 70m$) to $300 \times 700$ grids.



\begin{figure}
\centering
\includegraphics[width=\columnwidth]{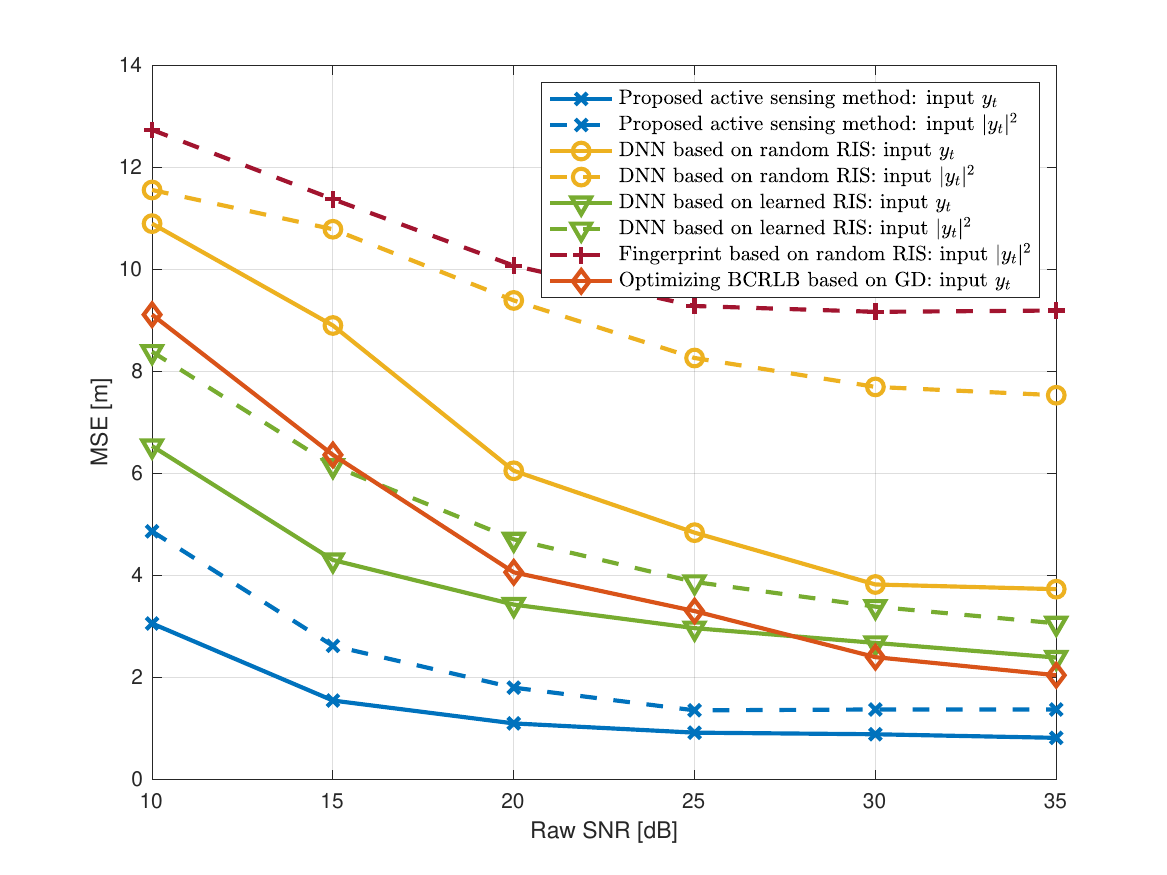}%
\caption{Localization accuracy vs. raw SNR in single-RIS SISO network with $N=64$, $T=6$.}%
\label{fig.loc_ris_mse_vs_snr}%
\end{figure}

\subsection{Simulation Results} \label{SISOsimulationresult}

We first examine the localization performance of the proposed method under different raw SNR, i.e., $P_u = 10^{{\rm SNR}/10}$mW, when $T = 6$.  
In Fig.~\ref{fig.loc_ris_mse_vs_snr}, whether the input is the pilot symbols or RSS value, 
the proposed adaptive RIS design is seen to have the smallest localization MSE across different SNRs as compared to other benchmarks with non-adaptive RIS design. 
This implies that the proposed algorithm effectively utilizes the current and historical measurements to design a more suitable RIS configuration for future time frames to minimize localization error. 

We point out that the fingerprinting-based approach along with wKNN classifier is seen to experience poor performance in this simulation setup. This is due to the randomness in the NLoS Rician channel model which adversely influences the fingerprint matching accuracy. The random and non-adaptive uplink RIS configuration used in the fingerprinting-based approach also contributes to the performance gap from methods with optimized uplink RIS configuration.

We also point out that BCRLB-minimization-based adaptive RIS design does not perform as well as the data-driven approach. The BCRLB-based method does not serve as an optimal RIS design 
for the following reasons. First,
an active sensing solution to (\ref{prob_formulation}) should design a series of sensing configuration jointly across $t$ to minimize the localization MSE, i.e., $\{{\mathcal{G}}^{(t)}(\cdot)\}_{t=0}^{T-1}$ as in (\ref{prob_formulation}). However, 
the BCRLB-based approach greedily treats the active sensing problem by designing individual ${\mathcal{G}}^{(t)}(\cdot)$ at each time frame. 
Secondly, the BCRLB can be a loose lower bound of the MSE, especially when SNR is low and/or the number of observations is limited\cite{crlbbound,crlbbound2,crlbbound3}. 
Finally, the optimization of BCRLB is a nonconvex optimization problem; it is difficult to find its true optimal solution. The designed RIS reflection coefficients are by no means optimal. 

\begin{figure}
\centering
\includegraphics[width=\columnwidth]{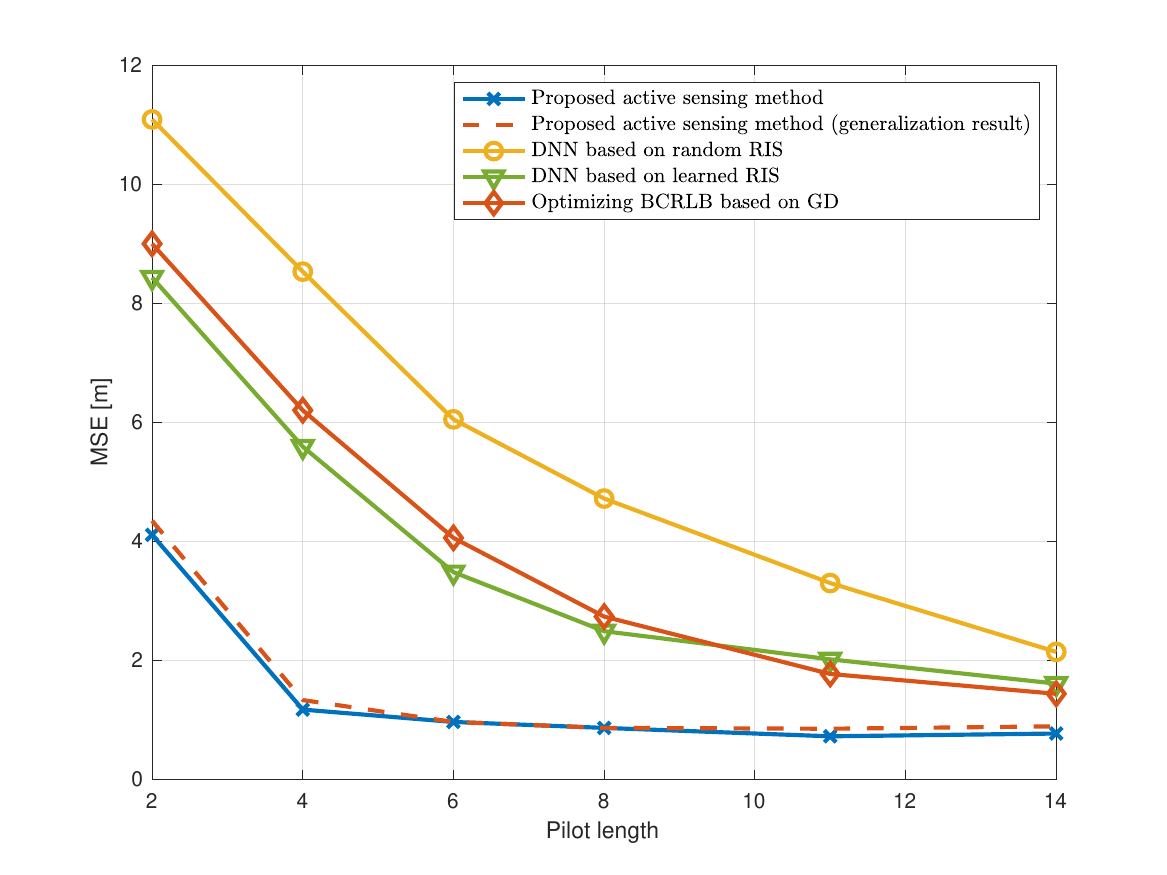}%
\caption{Localization accuracy vs. pilot length in single-RIS SISO network with $N=64$, SNR$=20$dB.}%
\label{fig.loc_ris_mse_vs_t}%
\end{figure}

\begin{figure}
\centering
\includegraphics[width=\columnwidth]{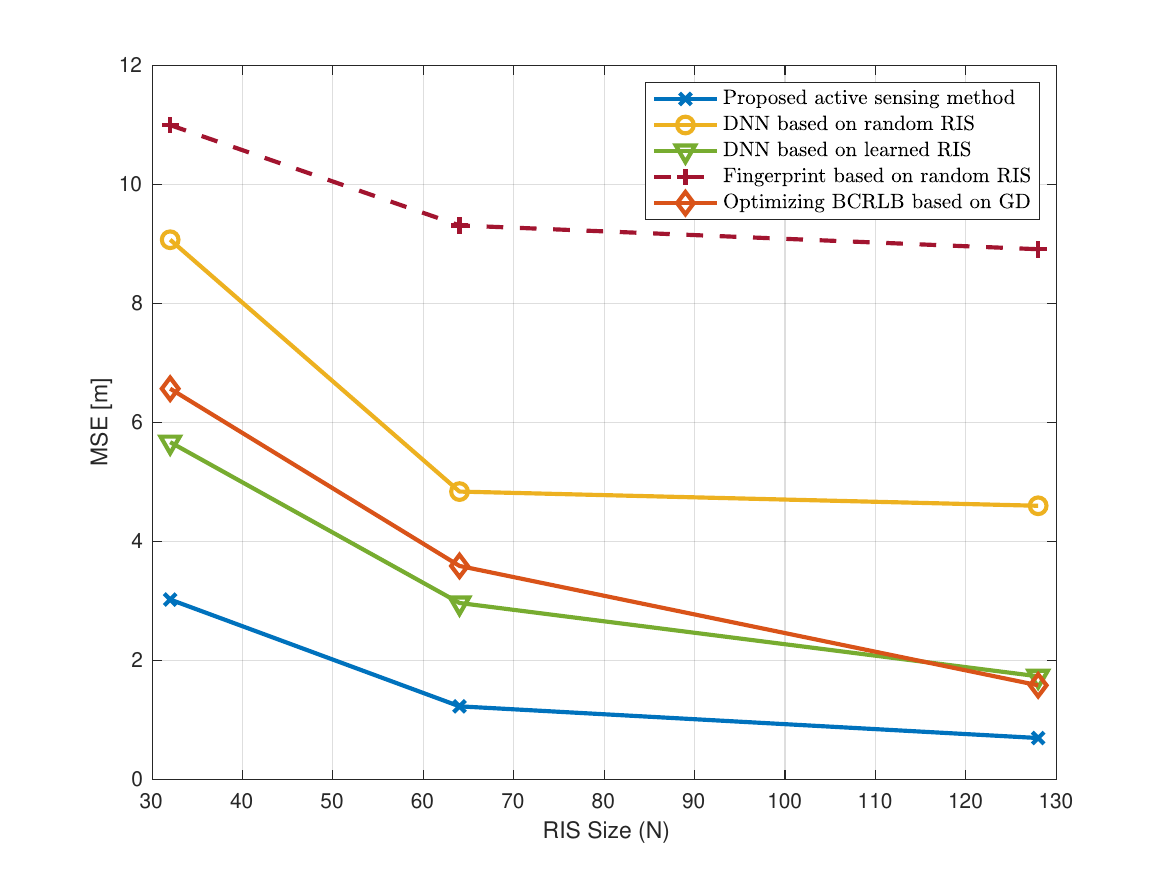}%
\caption{Localization accuracy vs. RIS size in single-RIS SISO network with SNR$=25$dB, $T=6$.}
    \label{fig.rissize}%
\end{figure}

We next examine the localization performance with varying numbers of time frames. From Fig.~\ref{fig.loc_ris_mse_vs_t}, we observe that the proposed algorithm demonstrates a consistently better performance over other benchmarks. For example, the proposed active sensing scheme with $4$ pilots can already achieve better localization performance than other baseline schemes with $14$ pilots. 
Further, the generalization curve shows that it is possible to train the model for $T=14$ and generalize to earlier stopping times with virtually no loss of performance.

We finally examine the localization performance with different sizes of RIS. From Fig.~\ref{fig.rissize}, we observe that as the size of the RIS grows larger, the associated localization accuracy increases. The performance gain is expected as a larger RIS provides richer angular information about the UE to the BS. 
The proposed learning-based approach is scalable with different sizes of RIS. To localize with different sizes of RIS, the only adjustment to the neural network architecture needed is to change the output dimension of the $(L+1)$-layer neural network at the output of the LSTM cells.

\begin{figure}[!t]
  \centering
  \begin{subfigure}{.33\columnwidth}
    \centering
    \includegraphics[width=\linewidth]{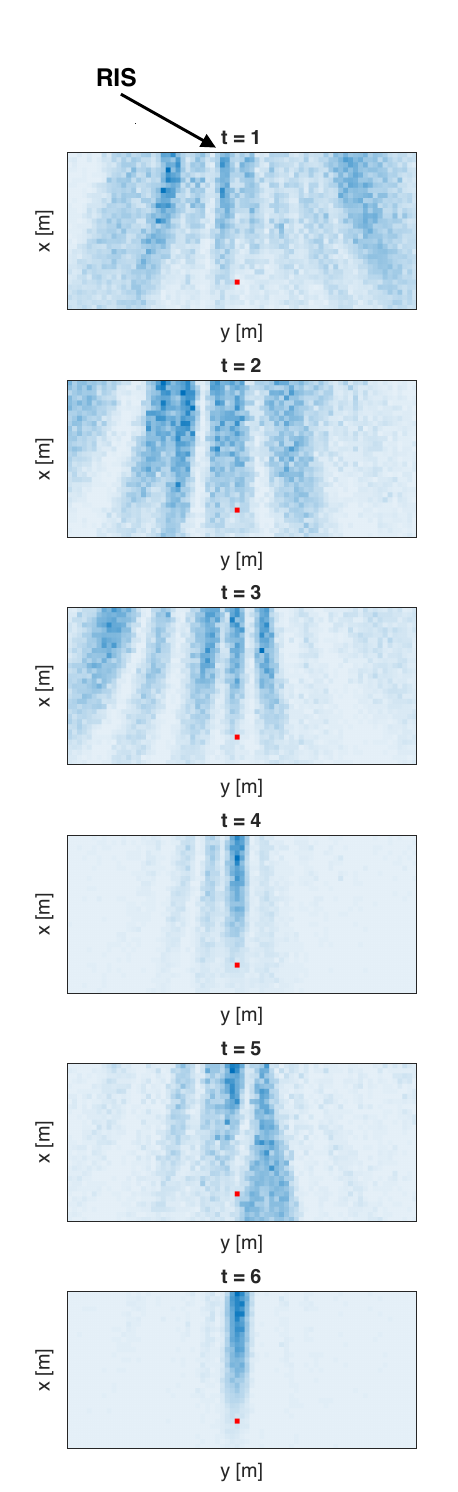}
\caption{The proposed method.}
    \label{loc_ris_interp_6_rnn}
  \end{subfigure}%
  \hfill
  \begin{subfigure}{.33\columnwidth}
    \centering
    \includegraphics[width=\linewidth]{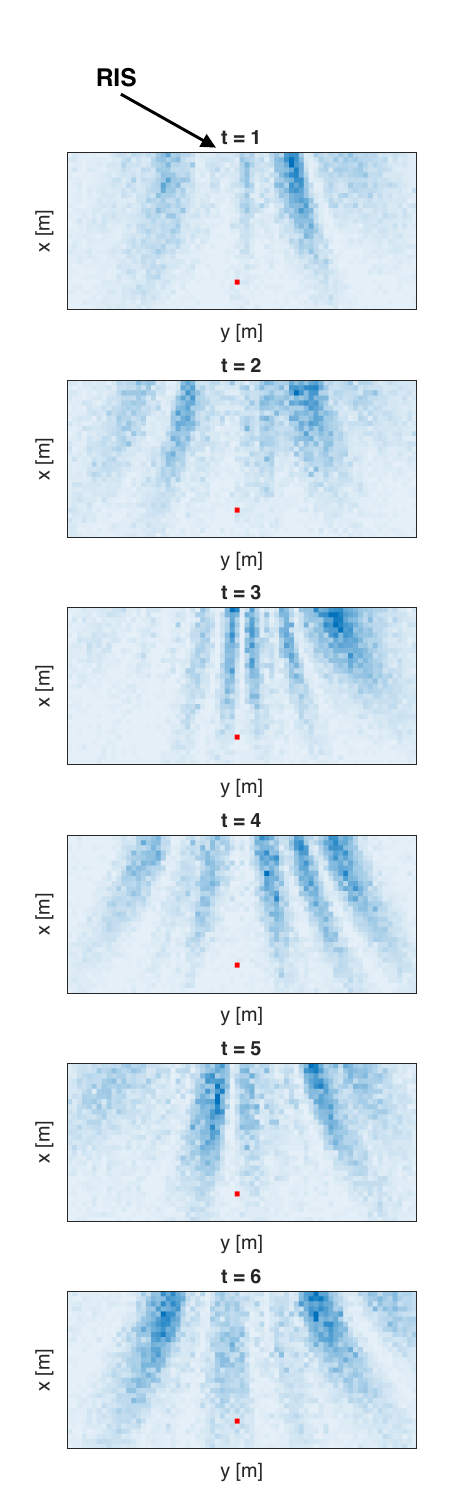}
\caption{Learned RIS (non-active).}%
    \label{loc_ris_interp_6_dnn}
  \end{subfigure}%
  \hfill
   \begin{subfigure}{.33\columnwidth}
    \centering
    \includegraphics[width=\linewidth]{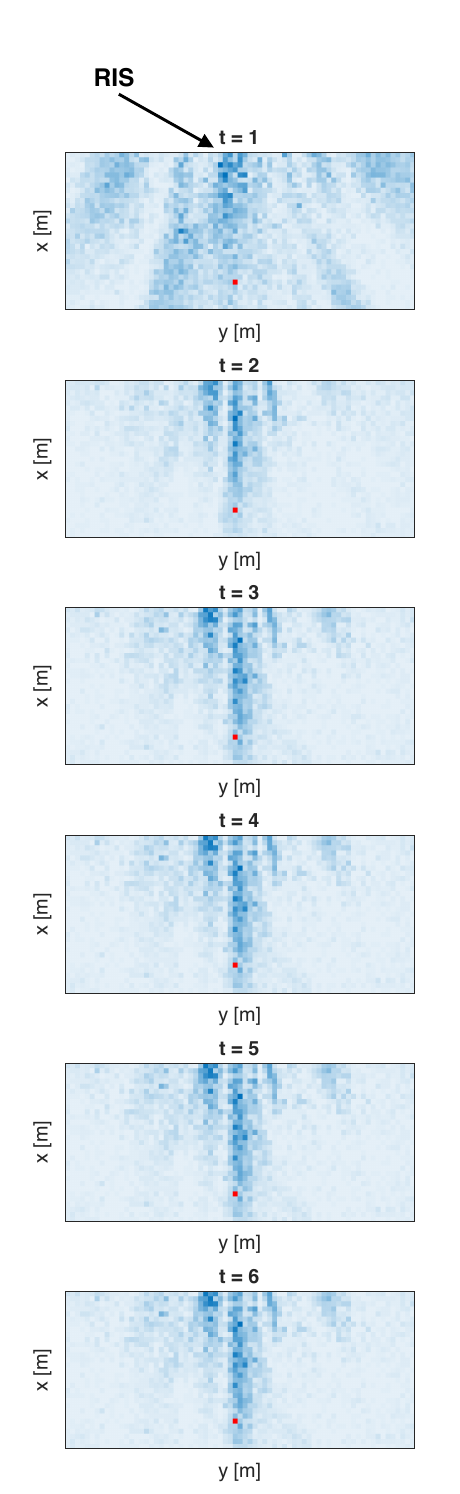}
\caption{BCRLB (active).}%
    \label{loc_ris_crlb}
  \end{subfigure}%
  \hfill
\caption{RSS distribution from time frame $1$ to $6$, $T=6$, SNR = $25$dB, single-RIS SISO network.}
    \label{loc_ris_interpretation}
\end{figure}

\subsection{Interpretation of the Solution Learned by Active Sensing}

We visually interpret the RIS design obtained from the proposed LSTM-based
network. Here, we test the neural network for a user with an arbitrary location and use the RSS
distribution (or radio map) to illustrate the beamforming pattern
produced by the adaptively designed RIS coefficients.  To do so, at each time
frame, we record the designed RIS configuration and plot the RSS obtained at
every $1m\times1m$ block in the area of interest across the $x-y$ plane 
as shown in Fig.~\ref{loc_ris_interpretation}. 
The red dot denotes the true UE position.
From Fig.~\ref{loc_ris_interpretation}(\subref{loc_ris_interp_6_rnn}), 
it is clear that the RIS is probing broader beams at the first several time frames to search for the user, then gradually focusing the beam towards the UE 
as $t$ increases. This implies that the proposed neural network is indeed designing meaningful RIS configurations based on the measurements. 
It is also worth noting that in Fig.~\ref{loc_ris_interpretation}(\subref{loc_ris_interp_6_rnn}), the beam is fixed towards the user at $t=4$, yet the beam widens in $t=5$. The LSTM network purposely comes up with such a non-greedy pattern to explore alternative directions. This is due to the fact that the sequence of RIS patterns is designed jointly across the $T$ sensing stages in order to produce an accurate estimation at the end. 
In comparison, the non-active scheme with RIS configurations learned from the channel statistics as shown in Fig.~\ref{loc_ris_interpretation}(\subref{loc_ris_interp_6_dnn}) 
does not exhibit beampatterns that focus over $t$. We also note that the radio map from BCRLB-based solution  Fig.~\ref{loc_ris_interpretation}(\subref{loc_ris_crlb}) also demonstrates focusing beampatterns but the beam towards the UE is not as strong and has large sidelobes. 


\subsection{Complexity Analysis}

The complexity scaling of the proposed learning-based approach is ${O}\left[ T \left( D_iZ_i + \Omega + \zeta(\sum_k N_k + M)  + D_oZ_o \right)  \right]  $, where $Z_i$ and $Z_o$ denote the dimension of the input layer and output layer ($\bm{\ell}_p$), respectively, $\Omega$ denotes the computational complexity of LSTM inference operation for one sensing stage, $\zeta$ denotes the computational complexity of the $(L+1)$-layer neural network to design the sensing vectors, $D_i$ and $D_o$ denote the dimension of input feature, e.g., pilots, and output feature respectively. Recall that $T$ denotes the number of sensing stages, $M$ denotes the number of antenna at the BS beamformer and $N_k$ denotes the number of reflective elements at the $k$-th RIS. 
It is important to note that both the training and inference processes are highly parallelizable using modern graphic process units (GPUs), so that in practice the proposed approach can be executed very efficiently.

\section{Numerical Results for Multi-RIS MISO Network}
\label{sec.numerical_irs_multi}

%



\begin{figure*}
\centering
\captionsetup{justification=centering}
\begin{subfigure}{\columnwidth}
\includegraphics[width=\columnwidth]{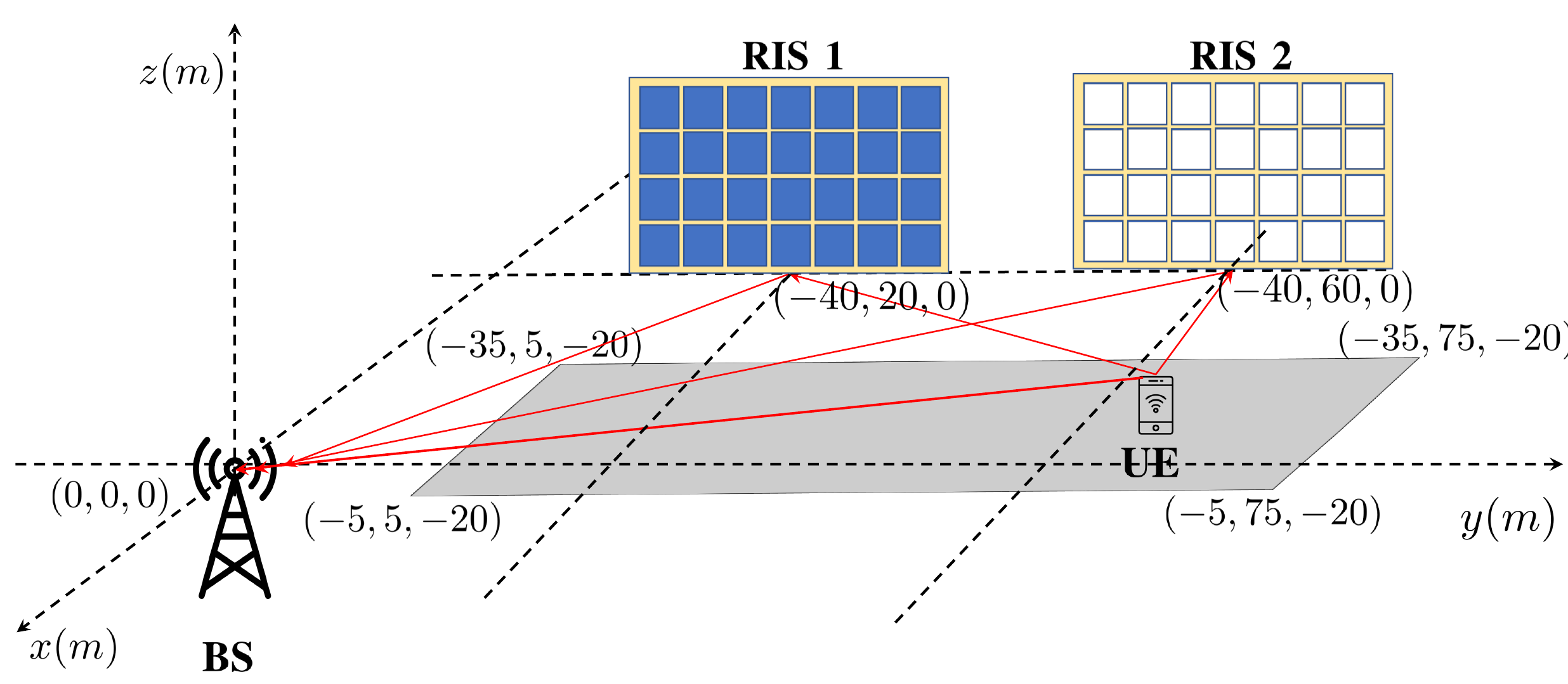}%
\caption{One-BS two-RIS network.}%
\label{fig.tworis}%
\end{subfigure}
\begin{subfigure}{\columnwidth}
\includegraphics[width=\columnwidth]{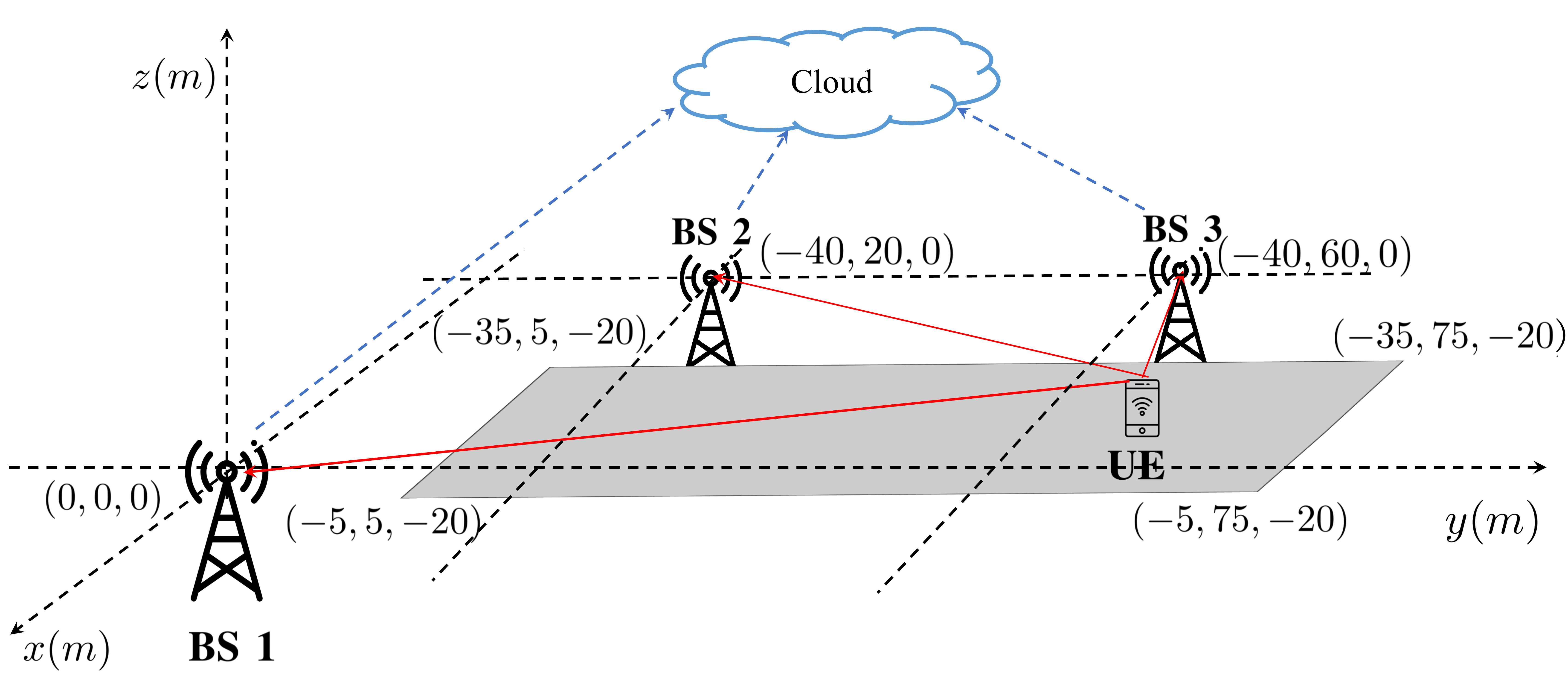}%
\caption{Three-BS network.}%
\label{fig.threebs}%
\end{subfigure}
\caption{Simulation environment for localization in a MISO Network.}
\label{label:simsetupmisonetwork}
\end{figure*}

In this section, we extend the proposed algorithm to multi-RIS network. 
Additional RISs enable better localization accuracy than the single-RIS setup shown in Section \ref{sec.numerical_irs_single}, because: \lowerromannumeral{1}) additional RISs provide new anchor points in the service area in which additional distance and angular information of the UE can be
deduced from the received measurements; \lowerromannumeral{2}) additional RISs enhance the coverage by reducing the distance between the RIS and the UE, which in turn reduces pathloss, and \lowerromannumeral{3}) multiple RISs enable constructive interference between the reflected signals which allows stronger beam focusing to the UE.

One of the key advantages of using RISs for localization is that RISs essentially
act as virtual anchors, thus allowing localization without multiple APs. 
As the RISs can be easily mounted on the ceilings and on the walls, this provides a 
cost-efficient and power-efficient way to provide localization services, especially
in large indoor areas such as industrial plants or warehouse settings with autonomous
vehicle and robot operations. 


Traditional localization requires triangulation with multiple anchors (BS or AP), each equipped
with individual signal processing units and RF chains. In this section, we show
that passive anchors, i.e., RISs, can be as effective as BS-based anchors. 
The advantage of such a hybrid active and passive anchor
trilateration setup is apparent as the RIS is cheaper to manufacture and to
maintain. They can be installed at many more sites. 
The authors of \cite{liangliu} conclude that two BS and one RIS are sufficient
to achieve trilateration. 
In this section, we provide evidence that a single BS with multiple RISs can already achieve excellent localization accuracy as compared to a conventional triangulation system---if the RISs are configured in an active manner. 





\subsection{System Setup}

We consider localization experiment setups with multiple RISs as shown 
in Fig.~\ref{label:simsetupmisonetwork}(\subref{fig.tworis})
as compared to the setup with multiple BSs as shown in 
Fig.~\ref{label:simsetupmisonetwork}(\subref{fig.threebs}).
As before, we assume that the BSs are equipped
with multiple antennas with analog sensing beamformers that can be designed actively. 
Each BS is equipped with a single RF chain. 

The design methodology of the multiple-RIS system is as described earlier in
the paper.  In the case of the multiple-BS system, we assume BS cooperation, i.e., 
a cloud center has access to the received pilots from all BSs.  
We generalize the following methodologies for configuring the sensing vector at the BSs: 
\subsubsection{Active sensing using LSTM}
The sequence of beamforming vectors is active over $T$ stages. At each time frame, the existing measurements at all the BSs are processed collectively in a central cloud to design the next set of beamforming vectors.  

\subsubsection{DNN with learned sensing vectors}
    The sequence of $T$ sets of sensing vectors is non-adaptive. Here, the sequence of sensing vectors is learned from training data, but is not adaptive as a function of previous measurements. All received measurements at the BSs are processed collectively via the cloud to determine the unknown UE position.


\subsection{Simulation Environment}
For the multi-RIS-assisted system, the BS with $M=8$ antennas and a single RF chain is placed at the origin, and $K=2$ RISs, each with $8 \times 8$ reflective elements, are deployed in the network to aid the UE localization as shown in Fig.~\ref{label:simsetupmisonetwork}(\subref{fig.tworis}). The two RISs are placed at $\bm{p}_1^{\rm RIS} = (-40m,20m,0m)$ and $\bm{p}_2^{\rm RIS} = (-40m,60m,0m)$. 

For the multi-BS system, we deploy three BSs at $\bm{p}_1^{\rm BS} = (0m,0m,0m)$, $\bm{p}_2^{\rm BS} = (-40m,20m,0m)$ and $\bm{p}_3^{\rm BS} = (-40m,60m,0m)$ respectively, as shown in Fig.~\ref{label:simsetupmisonetwork}(\subref{fig.threebs}). Each BS is equipped with $M=8$ antennas and a single RF chain. 

Same as the simulation environment in Section \ref{siso_singleris_environment}, the unknown user locations $\bm{p}$ are uniformly generated within a rectangular area on the $x$-$y$ plane $(-20m\pm15m, 40m\pm35m, -20m)$, as shown in Fig.~\ref{label:simsetupmisonetwork}. 
The Rician factor $\epsilon$ is set to $10$, and $ {2\pi d_{R}}/{\lambda_c} = 1$ without loss of generality. The path loss models of the direct and reflected paths are $32.6+36.7\log(d_1)$ and $30+22\log(d_2)$, respectively, where $d_1$ and $d_2$ denote the corresponding link distance. The bandwidth is $10$ MHz with a background noise of $-170$ dBm/Hz.

\begin{figure}
\centering
\includegraphics[width=\columnwidth]{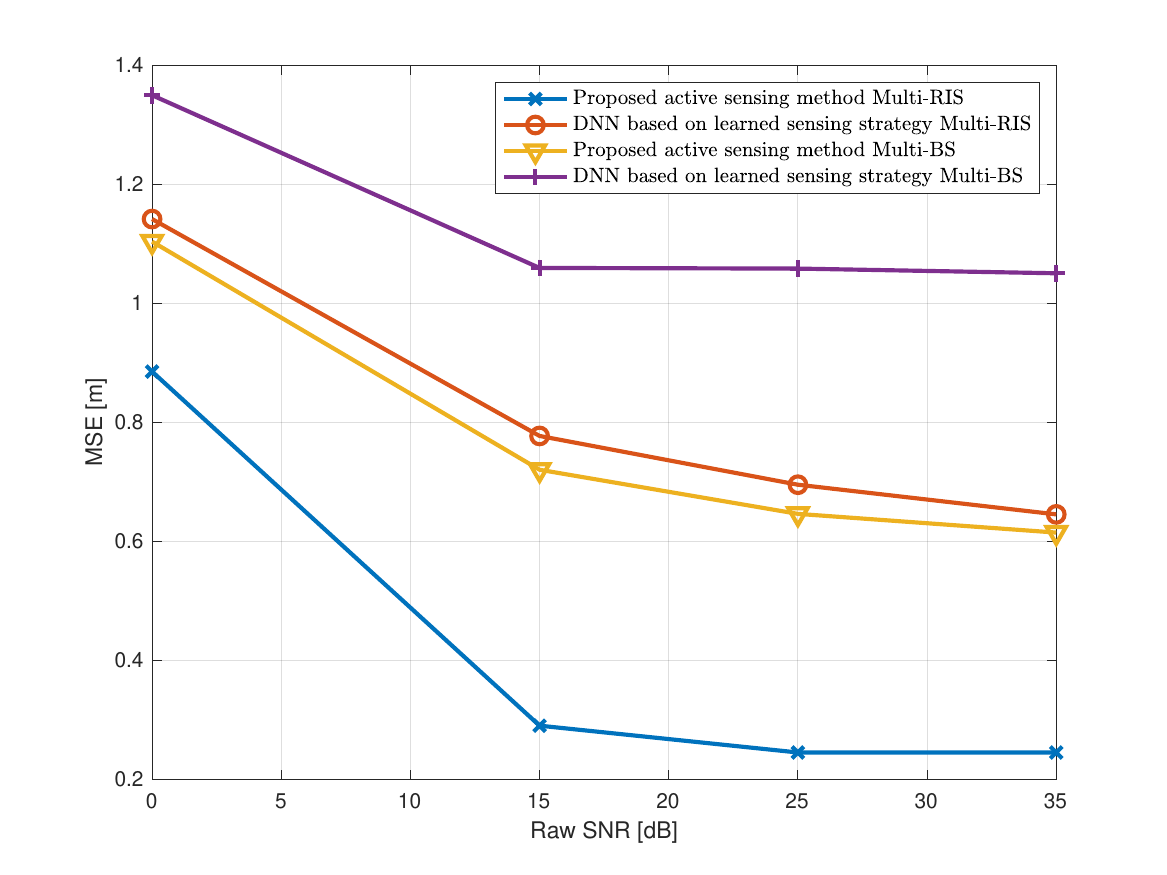}%
\caption{Performance comparison between 3BS system and 2RIS + 1BS system with  $T=6$.}%
\label{fig.3BSvs2RIS}%
\end{figure}

\subsection{Localization Performance with LoS Direct Channels}

The localization performances of multi-RIS-assisted system and multi-BS system with respect to raw SNR are shown in Fig.~\ref{fig.3BSvs2RIS}.  
Here, we assume that LoS channels exist between the BSs, the RISs, and the users.

Upon close examination of Fig.~\ref{fig.3BSvs2RIS}, in both the multi-RIS-assisted system and the multi-BS system, the proposed active sensing localization algorithm is seen to have a higher localization accuracy as compared to the non-active sensing benchmark, i.e., learned sensing algorithm. This result is consistent with the conclusion in single-RIS SISO setting from Section \ref{SISOsimulationresult}, which implies that the proposed active sensing algorithm is effective in utilizing the current and historical measurements to design a more suitable sensing vector for future time frames to minimize the localization error. 

We point out that it is remarkable that the multi-RIS-assisted system results in lower localization error as compared to the multi-BS system under the same localization algorithms. 
We emphasize that the RIS typically has more reflective elements as compared to the number of antennas at the BS.
The resulting performance advantage is because of the larger beamforming gain and the higher angular resolution due to the large number of reflective elements at the RISs. 
But it is remarkable that the advantage holds despite the larger path loss in the reflected paths!

Lastly, we compare the performance of multi-RIS-assisted MISO system from Fig.~\ref{fig.3BSvs2RIS} against the single-RIS-assisted SISO system from Fig.~\ref{fig.loc_ris_mse_vs_snr}. By comparing the proposed active sensing algorithm and learned sensing algorithm from the two figures, 
additional improvement in localization accuracy is observed in the multi-RIS-assisted system. This implies that the additional RIS and the additional antennas at the BS provide extra beamforming capabilities which enable better localization. 

\begin{figure}[!t]
  \centering
  \begin{subfigure}{.33\columnwidth}
    \centering
    \includegraphics[width=\linewidth]{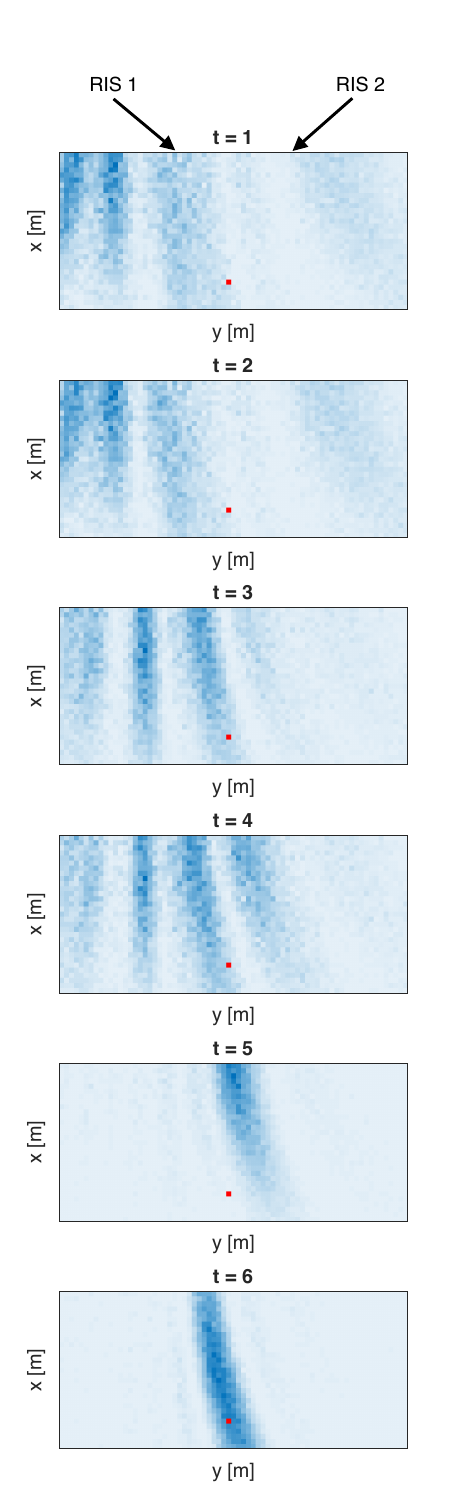}
\caption{RIS 1 radio map.}
    \label{loc_ris_interp_6_rnn_ris1}
  \end{subfigure}%
  \hfill
  \begin{subfigure}{.33\columnwidth}
    \centering
    \includegraphics[width=\linewidth]{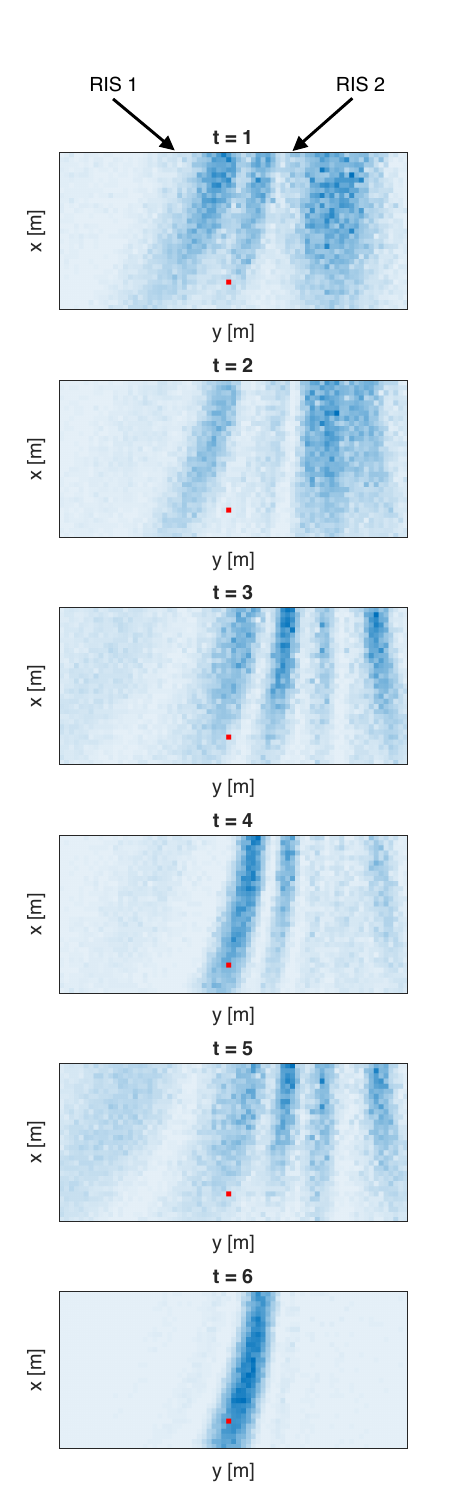}
\caption{RIS 2 radio map.}%
    \label{loc_ris_interp_6_rnn_ris2}
  \end{subfigure}%
  \hfill
   \begin{subfigure}{.33\columnwidth}
    \centering
    \includegraphics[width=\linewidth]{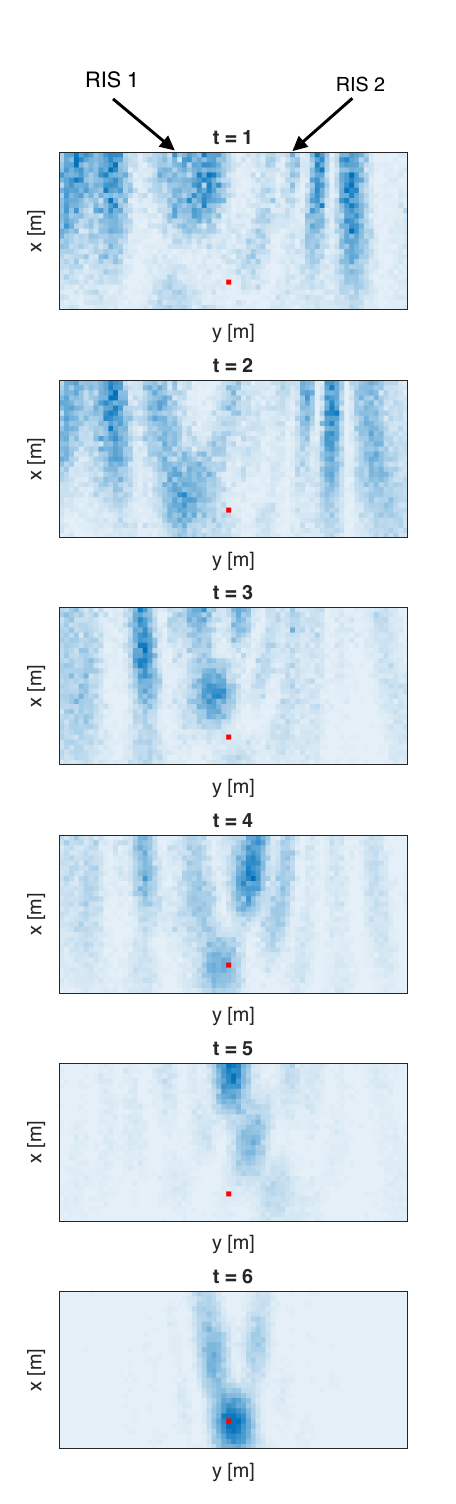}
\caption{RIS 1 and 2 radio map.}
    \label{loc_ris_interp_6_rnn_bothris}
  \end{subfigure}%
  \hfill
  \captionsetup{justification=centering}
\caption{The RSS distribution of the proposed active sensing method from time frame $1$ to $6$ with raw SNR = $35$dB in a multi-RIS MISO network.}
    \label{loc_2ris_interpretation}
\end{figure}

\subsection{Interpretation of the Solution Learned by Active Sensing}

We interpret the sensing vectors obtained from the proposed LSTM-based network. 
Here, we test the neural network for a user with an arbitrary location and use the RSS
distribution (or radio map) as a means to illustrate the beamforming patterns
produced by the adaptively designed RIS configurations.  
To do so, at each time frame, we separately record the beamforming and the reflection patterns corresponding to the designed configurations of RIS 1, and the designed configurations of RIS 2, together with the designed BS beamformer. 
Next, we plot the three RSS distributions obtained at
every $1m\times1m$ block in the service area across the $x-y$ plane 
as shown in Fig.~\ref{loc_2ris_interpretation}. The three radio maps correspond to: 1) RSS distribution produced by RIS 1 and the beamforming vector; 2) RSS distribution produced by RIS 2 and the beamforming vector; and 3) RSS distribution produced by RIS 1, RIS 2 and the beamforming vector together.
Here, the red dot denotes the true UE position.
From Fig.~\ref{loc_2ris_interpretation}(\subref{loc_ris_interp_6_rnn_ris1}), 
it is clear that RIS 1 is probing broader beams at the first several time frames to search for the user, then gradually focusing the beam towards the UE 
as $t$ increases. It is also clear that the design of RIS 2 adopts the same strategy from Fig.~\ref{loc_2ris_interpretation}(\subref{loc_ris_interp_6_rnn_ris2}). 
The combined beampattern from RIS 1 and RIS 2 shown in Fig.~\ref{loc_2ris_interpretation}(\subref{loc_ris_interp_6_rnn_bothris}) indicates a constructive interference within the proximity of the UE and destructive interference outside of the proximity as $t$ increases. 
This type of interference pattern is the reason for accurate localization. 
It illustrates how the proposed neural network indeed designs meaningful and interpretable 
sensing vectors to narrow down the user location based on the prior observations. 



\subsection{Localization Performance with NLoS Direct Channels}

\begin{figure}
\centering
\includegraphics[width=\columnwidth]{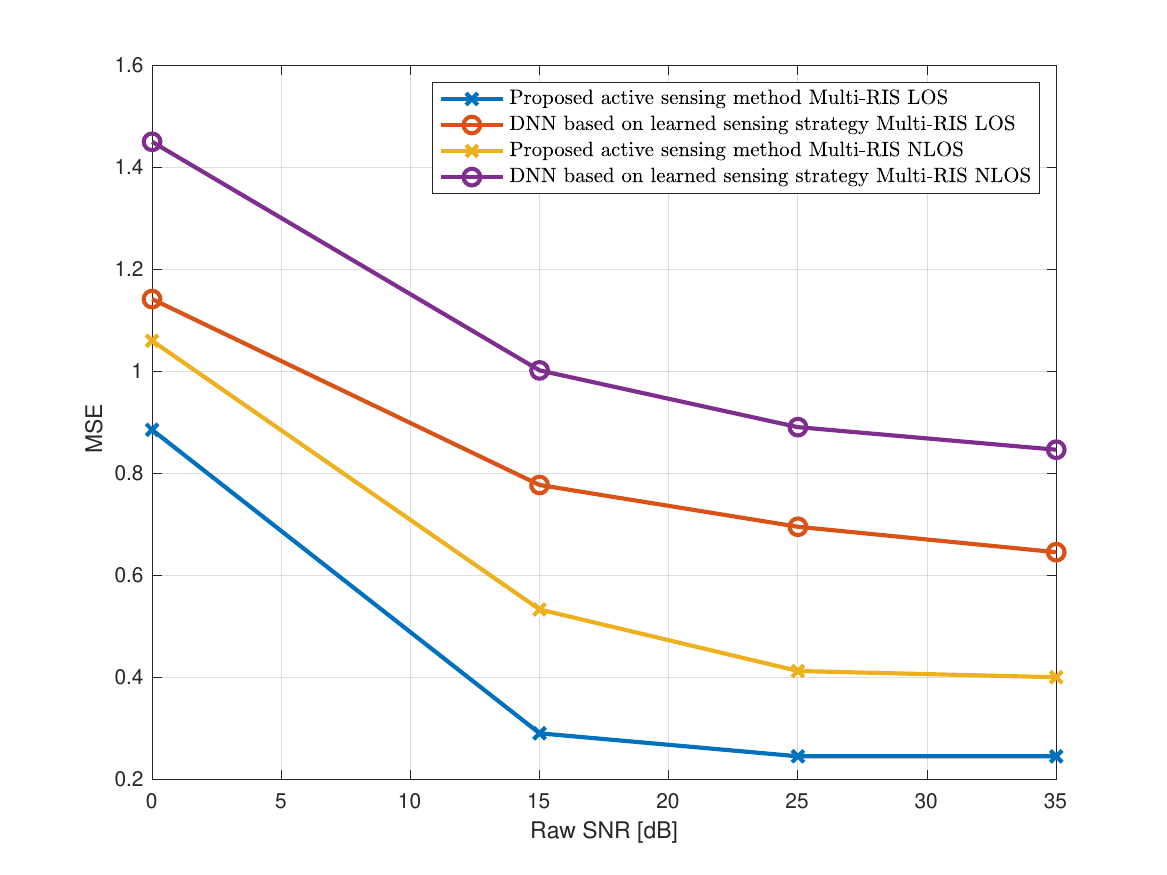}%
\caption{Performance comparison between system with LoS direct channels and system with NLoS direct channels for $T=6$ in the multi-RIS MISO network.}%
\label{fig.3BSvs2RIS_NLOS}%
\end{figure}

An attractive feature of the RIS is that it can provide reliable coverage to users even when there is no direct LoS between the BS and the UE. To illustrate this point,
we examine a deployment scenario when LoS exists between the BS and the RISs, between the RISs and the UE, but not between the BS and the UE. 
As shown in Fig.~\ref{fig.3BSvs2RIS_NLOS}, excellent localization performance can already be obtained based on the reflected paths, without LoS in the direct path.

As also shown in Fig.~\ref{fig.3BSvs2RIS_NLOS}, if there is a direct LoS between the BS and the UE, then the localization performance would further improve. The performance gain is, however, not as drastic as one would expect. This is because the reflected paths already contain a lot of information in the angular domain---due to the large number of reflective elements at the two RISs.


\section{conclusions}
\label{sec.conclusions}

This paper shows that active sensing can significantly improve the performance of
an uplink RIS-assisted localization task in which the BS adaptively designs the 
sequence of sensing vectors including the beamformers and the RIS reflection coefficients based on the received pilots from the user to enhance the localization accuracy.  
By employing an LSTM-based neural network, the proposed solution
can effectively utilize historical measurements to design the RIS configurations and the beamforming vectors
for subsequent measurements in a codebook-free fashion for the purpose of
minimizing localization error. Numerical results indicate a significantly lower localization 
error as compared to the existing benchmarks. The proposed solution demonstrates interpretable results and is generalizable to early stopping in the sequence of sensing stages. The paper also shows that passive RISs can effectively replace active APs as anchors for achieving accurate localization performance. This shows the significant advantage of deploying RISs for location services.

\bibliography{references}

\begin{thebibliography}{10}
\providecommand{\url}[1]{#1}
\csname url@samestyle\endcsname
\providecommand{\newblock}{\relax}
\providecommand{\bibinfo}[2]{#2}
\providecommand{\BIBentrySTDinterwordspacing}{\spaceskip=0pt\relax}
\providecommand{\BIBentryALTinterwordstretchfactor}{4}
\providecommand{\BIBentryALTinterwordspacing}{\spaceskip=\fontdimen2\font plus
\BIBentryALTinterwordstretchfactor\fontdimen3\font minus
  \fontdimen4\font\relax}
\providecommand{\BIBforeignlanguage}[2]{{%
\expandafter\ifx\csname l@#1\endcsname\relax
\typeout{** WARNING: IEEEtran.bst: No hyphenation pattern has been}%
\typeout{** loaded for the language `#1'. Using the pattern for}%
\typeout{** the default language instead.}%
\else
\language=\csname l@#1\endcsname
\fi
#2}}
\providecommand{\BIBdecl}{\relax}
\BIBdecl

\bibitem{active2023icc}
Z.~Zhang, T.~Jiang, and W.~Yu, ``Active sensing for localization with
  reconfigurable intelligent surface,'' in \emph{Proc. IEEE Int. Conf. Commun.
  (ICC)}, Jun. 2023, pp. 4261--4266.

\bibitem{6924849}
R.~Di~Taranto, S.~Muppirisetty, R.~Raulefs, D.~Slock, T.~Svensson, and
  H.~Wymeersch, ``Location-aware communications for {5G} networks: How location
  information can improve scalability, latency, and robustness of {5G},''
  \emph{IEEE Signal Process. Mag.}, vol.~31, no.~6, pp. 102--112, Oct. 2014.

\bibitem{6gwhitepaper}
\BIBentryALTinterwordspacing
A.~Bourdoux \emph{et~al.}, ``{6G} white paper on localization and sensing,''
  2020. [Online]. Available: \url{https://arxiv.org/abs/2006.01779}
\BIBentrySTDinterwordspacing

\bibitem{Palaskar2014WiFiIP}
P.~Palaskar, R.~Palkar, and M.~Tawari, ``{Wi-Fi} indoor positioning system
  based on {RSSI} measurements from {Wi-Fi} access points -a tri-lateration
  approach,'' \emph{Int.\ J.\ Scientific Eng. Res.}, vol.~5, no.~4, pp.
  1234--1238, Apr. 2014.

\bibitem{4796924}
Y.~Gu, A.~Lo, and I.~Niemegeers, ``A survey of indoor positioning systems for
  wireless personal networks,'' \emph{IEEE Commun. Surveys Tutorials}, vol.~11,
  no.~1, pp. 13--32, Mar. 2009.

\bibitem{aoa_localization}
E.~Y. Menta, N.~Malm, R.~Jäntti, K.~Ruttik, M.~Costa, and K.~Leppänen, ``On
  the performance of {AoA}–based localization in {5G} ultra–dense
  networks,'' \emph{IEEE Access}, vol.~7, pp. 33\,870--33\,880, Mar. 2019.

\bibitem{aoa_localization2}
J.~Xu, M.~Ma, and C.~L. Law, ``{AOA} cooperative position localization,'' in
  \emph{Proc.\ IEEE Global Commun. (Globecom)}, Nov. 2008, pp. 1--5.

\bibitem{nguyen}
C.~L. Nguyen, O.~Georgiou, G.~Gradoni, and M.~Di~Renzo, ``Wireless
  fingerprinting localization in smart environments using reconfigurable
  intelligent surfaces,'' \emph{IEEE Access}, vol.~PP, pp. 1--1, Sept. 2021.

\bibitem{bible}
M.~Di~Renzo, A.~Zappone, M.~Debbah, M.-S. Alouini, C.~Yuen, J.~de~Rosny, and
  S.~Tretyakov, ``Smart radio environments empowered by reconfigurable
  intelligent surfaces: How it works, state of research, and the road ahead,''
  \emph{IEEE J.\ Select.\ Areas Commun.}, vol.~38, no.~11, pp. 2450--2525, Jul.
  2020.

\bibitem{basar2019wireless}
E.~{Basar}, M.~{Di Renzo}, J.~{de Rosny}, M.~{Debbah}, M.-S. {Alouini}, and
  R.~{Zhang}, ``Wireless communications through reconfigurable intelligent
  surfaces,'' \emph{IEEE Access}, vol.~7, pp. 116\,753--116\,773, Aug. 2019.

\bibitem{4fuliu}
F.~Liu, O.~Tsilipakos, A.~Pitilakis, A.~C. Tasolamprou, M.~S. Mirmoosa, N.~V.
  Kantartzis, D.-H. Kwon, M.~Kafesaki, C.~M. Soukoulis, and S.~A. Tretyakov,
  ``Intelligent metasurfaces with continuously tunable local surface impedance
  for multiple reconfigurable functions,'' \emph{Phys.\ Rev.\ A Gen.\ Phys.},
  vol.~11, no.~4, pp. 2331--7019, Apr. 2019.

\bibitem{coding_metasurface}
L.~{Li}, T.~J. {Cui}, W.~{Ji}, S.~{Liu}, J.~{Ding}, X.~{Wan}, Y.~B. {Li},
  M.~{Jiang}, C.-W. {Qiu}, and S.~{Zhang}, ``Electromagnetic reprogrammable
  coding-metasurface holograms,'' \emph{Nature Commun.}, vol.~8, no.~1, p. 197,
  Aug. 2017.

\bibitem{panpan}
C.~Pan \emph{et~al.}, ``An overview of signal processing techniques for
  {RIS}/{IRS}-aided wireless systems,'' \emph{IEEE J.\ Sel.\ Topics Signal
  Process.}, vol.~16, no.~5, pp. 1--35, Aug. 2022.

\bibitem{hejiguang}
J.~He, H.~Wymeersch, T.~Sanguanpuak, O.~Silven, and M.~Juntti, ``Adaptive
  beamforming design for {mmWave} {RIS}-aided joint localization and
  communication,'' in \emph{Proc. IEEE Wireless Commun. Netw. Conf. Workshops
  (WCNCW)}, Apr. 2020, pp. 1--6.

\bibitem{liu2020reconfigurable}
Y.~Liu, E.~Liu, R.~Wang, and Y.~Geng, ``Reconfigurable intelligent surface
  aided wireless localization,'' in \emph{Proc. IEEE Int. Conf. Commun. (ICC)},
  Jun. 2021, pp. 1--6.

\bibitem{siso}
K.~Keykhosravi, M.~F. Keskin, G.~Seco-Granados, and H.~Wymeersch, ``{SISO}
  {RIS}-enabled joint {3D} downlink localization and synchronization,'' in
  \emph{Proc. IEEE Int. Conf. Commun. (ICC)}, Jun. 2021, pp. 1--6.

\bibitem{nearfield}
Z.~Abu-Shaban, K.~Keykhosravi, M.~F. Keskin, G.~C. Alexandropoulos,
  G.~Seco-Granados, and H.~Wymeersch, ``Near-field localization with a
  reconfigurable intelligent surface acting as lens,'' in \emph{Proc. IEEE Int.
  Conf. Commun. (ICC)}, Jun. 2021, pp. 1--6.

\bibitem{mimoofdm}
Y.~Lin, S.~Jin, M.~Matthaiou, and X.~You, ``Channel estimation and user
  localization for {IRS}-assisted {MIMO-OFDM} systems,'' \emph{IEEE Trans.\
  Wireless Commun.}, vol.~21, no.~4, pp. 2320--2335, Apr. 2022.

\bibitem{Ubiquitous}
H.~Zhang, H.~Zhang, B.~Di, K.~Bian, Z.~Han, and L.~Song, ``Towards ubiquitous
  positioning by leveraging reconfigurable intelligent surface,'' \emph{IEEE
  Commun.\ Lett.}, vol.~25, no.~1, pp. 284--288, Sept. 2021.

\bibitem{Ubiquitous2}
------, ``Metalocalization: Reconfigurable intelligent surface aided multi-user
  wireless indoor localization,'' \emph{IEEE Trans.\ Wireless Commun.},
  vol.~20, no.~12, pp. 7743--7757, Jun. 2021.

\bibitem{wang2021joint}
R.~Wang, Z.~Xing, E.~Liu, and J.~Wu, ``Joint localization and communication
  study for intelligent reflecting surface aided wireless communication
  system,'' \emph{IEEE Trans.\ Wireless Commun.}, vol.~71, no.~5, pp.
  3024--3042, Feb. 2023.

\bibitem{elzanaty2020reconfigurable}
A.~Elzanaty, A.~Guerra, F.~Guidi, and M.-S. Alouini, ``Reconfigurable
  intelligent surfaces for localization: Position and orientation error
  bounds,'' \emph{IEEE Trans.\ Signal Process.}, vol.~69, pp. 5386--5402, Aug.
  2021.

\bibitem{combination}
H.~Wymeersch and B.~Denis, ``Beyond {5G} wireless localization with
  reconfigurable intelligent surfaces,'' in \emph{Proc. IEEE Int. Conf. Commun.
  (ICC)}, Jun. 2020, pp. 1--6.

\bibitem{multiriscrlb}
Y.~Liu, S.~Hong, C.~Pan, Y.~Wang, Y.~Pan, and M.~Chen, ``Cramér-rao lower
  bound analysis of multiple-{RIS}-aided {mmWave} positioning systems,'' in
  \emph{Proc.\ IEEE Int.\ Symp.\ Pers.\ Indoor Mobile Radio Commun.\ (PIMRC)},
  Sept. 2022, pp. 1110--1115.

\bibitem{aoa_adaptive}
F.~Sohrabi, Z.~Chen, and W.~Yu, ``Deep active learning approach to adaptive
  beamforming for {mmWave} initial alignment,'' in \emph{Proc.\ IEEE Int.\
  Conf.\ Acoust.\ Speech, Signal Processing (ICASSP)}, Jun. 2021, pp.
  4940--4944.

\bibitem{sohrabi2021active}
F.~Sohrabi, T.~Jiang, W.~Cui, and W.~Yu, ``Active sensing for communications by
  learning,'' \emph{IEEE J.\ Select.\ Areas Commun.}, vol.~40, no.~6, pp.
  1780--1794, Jun. 2022.

\bibitem{javidi}
S.~Chiu, N.~Ronquillo, and T.~Javidi, ``Active learning and {CSI} acquisition
  for mmwave initial alignment,'' \emph{IEEE J.\ Select.\ Areas Commun.},
  vol.~37, no.~11, pp. 2474--2489, Nov. 2019.

\bibitem{aoa_adaptive_journal}
F.~Sohrabi, Z.~Chen, and W.~Yu, ``Deep active learning approach to adaptive
  beamforming for {mmWave} initial alignment,'' \emph{IEEE J.\ Select.\ Areas
  Commun.}, vol.~39, no.~8, pp. 2347--2360, Aug. 2021.

\bibitem{GRU}
A.~Sant, A.~Abdi, and J.~Soriaga, ``Deep sequential beamformer learning for
  multipath channels in mmwave communication systems,'' in \emph{Proc.\ IEEE
  Int.\ Conf.\ Acoust.\ Speech, Signal Processing (ICASSP)}, May 2022, pp.
  5198--5202.

\bibitem{hantracking}
H.~Han, T.~Jiang, and W.~Yu, ``Active beam tracking with reconfigurable
  intelligent surface,'' in \emph{Proc.\ IEEE Int.\ Conf.\ Acoust.\ Speech,
  Signal Processing (ICASSP)}, Jun. 2023, pp. 1--5.

\bibitem{LSTM}
S.~Hochreiter and J.~Schmidhuber, ``Long short-term memory,'' \emph{Neural
  Comput.}, vol.~9, no.~8, p. 1735–1780, Nov. 1997.

\bibitem{taojournal}
T.~{Jiang}, H.~V. {Cheng}, and W.~{Yu}, ``Learning to reflect and to beamform
  for intelligent reflecting surface with implicit channel estimation,''
  \emph{IEEE J.\ Select.\ Areas Commun.}, vol.~39, no.~7, pp. 1931 -- 1945,
  Jul. 2021.

\bibitem{multiAIRS}
T.~Zhou, K.~Xu, Z.~Shen, W.~Xie, D.~Zhang, and J.~Xu, ``{AoA}-based positioning
  for aerial intelligent reflecting surface-aided wireless communications: An
  angle-domain approach,'' \emph{IEEE Commun.\ Lett.}, vol.~11, no.~4, pp.
  761--765, Apr. 2022.

\bibitem{adaptiveMIMO}
W.~Huleihel, J.~Tabrikian, and R.~Shavit, ``Optimal adaptive waveform design
  for cognitive {MIMO} radar,'' \emph{IEEE Trans.\ Signal Processing}, vol.~61,
  no.~20, pp. 5075--5089, Jun. 2013.

\bibitem{adam}
\BIBentryALTinterwordspacing
D.~P. Kingma and J.~Ba, ``Adam: {A} method for stochastic optimization,'' in
  \emph{Proc. Int. Conf. Learn. Representations (ICLR)}, 2015. [Online].
  Available: \url{http://arxiv.org/abs/1412.6980}
\BIBentrySTDinterwordspacing

\bibitem{conditional_bcrlb}
L.~Zuo, R.~Niu, and P.~K. Varshney, ``Conditional posterior cramér–rao lower
  bounds for nonlinear sequential bayesian estimation,'' \emph{IEEE
  Transactions on Signal Processing}, vol.~59, no.~1, pp. 1--14, Sept. 2011.

\bibitem{hguo}
H.~{Guo}, Y.~{Liang}, J.~{Chen}, and E.~G. {Larsson}, ``Weighted sum-rate
  maximization for reconfigurable intelligent surface aided wireless
  networks,'' \emph{IEEE Trans.\ Wireless Commun.}, vol.~19, no.~5, pp.
  3064--3076, May 2020.

\bibitem{tensorflow}
M.~{Abadi} \emph{et~al.}, ``Tensorflow: A system for large-scale machine
  learning,'' in \emph{Proc. USENIX Conf. Operating Syst. Des. and
  Implementation (OSDI)}, 2016, p. 265–283.

\bibitem{crlbbound}
C.~Prévost, E.~Chaumette, K.~Usevich, D.~Brie, and P.~Comon, ``On cramér-rao
  lower bounds with random equality constraints,'' in \emph{Proc.\ IEEE Int.\
  Conf.\ Acoust.\ Speech, Signal Processing (ICASSP)}, May 2020, pp.
  5355--5359.

\bibitem{crlbbound2}
Y.~Noam and H.~Messer, ``Notes on the tightness of the hybrid cramÉr–rao
  lower bound,'' \emph{IEEE Trans.\ Signal Processing}, vol.~57, no.~6, pp.
  2074--2084, Feb. 2009.

\bibitem{crlbbound3}
A.~L. Matveyev, A.~B. Gershman, and J.~F. B\"{o}hme, ``On the direction
  estimation cram\'{e}r-rao bounds in the presence of uncorrelated unknown
  noise,'' \emph{Circuits Syst. Signal Process.}, vol.~18, no.~5, p. 479–487,
  Sept. 1999.

\bibitem{liangliu}
Q.~Wang, L.~Liu, S.~Zhang, and F.~C. Lau, ``Trilateration-based device-free
  sensing: Two base stations and one passive {IRS} are sufficient,'' in
  \emph{Proc.\ IEEE Global Commun. (Globecom)}, Dec. 2022, pp. 5613--5618.

\end{thebibliography}
\bibliographystyle{IEEEtran}

\end{document}